\documentclass[aps,prd,twocolumn,10pt]{revtex4-1}
\usepackage{xcolor}
\definecolor{Gray}{gray}{0.85}
\definecolor{LightCyan}{rgb}{0.88,1,1}
\usepackage[colorlinks,linkcolor={red},citecolor={blue}]{hyperref}
\usepackage{eurosym}
\usepackage{graphicx}
\usepackage{amsmath}
\usepackage{amssymb}
\usepackage{tensor}
\usepackage{amsfonts}
\usepackage{bm}
\usepackage{hhline}
\usepackage{float}

\usepackage{multirow}

\def\be{\begin{equation}}
	\def\ee{\end{equation}}
\def\bea{\begin{eqnarray}}
	\def\eea{\end{eqnarray}}

\newcommand{\f}[2]{\frac{#1}{#2}}

\begin{document}
\title{Cosmology in generalized hybrid metric-Palatini with matter-geometry coupling}
\author{Reza Jalali}
\email{s.jalali@email.kntu.ac.ir}
\author{Shahab Shahidi}
\email{s.shahidi@du.ac.ir}
\author{Mohammad Hossein Zhoolideh Haghighi}
\email{zhoolideh@kntu.ac.ir}
\affiliation{Department of Physics, K.N. Toosi University of Technology, Tehran, P.O. Box 15875-4416, Tehran, Iran.}
\affiliation{School of Physics, Damghan University, Damghan 36716-45667, Iran.}
\date{\today}

\begin{abstract}
Cosmological implications of a class of hybrid metric-Palatini gravity with a non-minimal matter-geometry coupling is considered. The theory contains a metric curvature tensor, together with a curvature tensor constructed from an independent affine connection. We will show that the model could be written as a bi-scalar-tensor gravity with a non-minimal coupling between matter sector and a scalar field. The theory will then be confronted with observational data from Cosmic Chronometers, BAO dataset from DESI and the Pantheon$^+$ dataset. We will show that the theory could be a good alternative to the $\Lambda$CDM model with the difference that the conservation of the baryonic matter sector holds only at the background level. The statefinder analysis will also be applied to the theory and it is observed that the DE behavior of the theory exhibits a quintessence to phantom transition occurs at redshifts around $z\approx0.86$. 
\end{abstract}
\maketitle

\section{Introduction}\label{secI}
Late time acceleration of the universe is one of the most interesting and also mysterious observations of the modern cosmology \cite{acceleration}. Observations of type Ia supernovae from Supernova Cosmology Project \cite{SCP} and High-Z Supernova Search Team \cite{HZ} proved that the universe experiences an accelerated expanding phase at the present time. This is also confirmed by BAO observations from different sources \cite{DESIDR2}. The accelerated expansion of the late time universe is associated with the presence of a well-known dark energy (DE) component with the property of having negative pressure, producing repulsive gravity. Among all ideas for the origin of DE, the simplest model is the cosmological constant $\Lambda$, which together with the Baryonic and dark matter components, form the well-know $\Lambda$CDM model \cite{LCDM}. This provides a non-dynamical explanation of the negative pressure part of the energy budget of our universe. Despite all the observational successes of the standard $\Lambda$CDM model, it suffers from a number of theoretical, observational and phenomenological issues that motivate cosmologists to search for an alternative. From the theoretical part, we face the cosmological constant problem which arises when one tries to explain the current observational value of the cosmological constant energy density $\rho_\Lambda\lesssim10^{-47}GeV$ from theoretical expectations of particle physics $\rho_c\lesssim10^{71}GeV$, differ from each other by more than $100$ orders of magnitude \cite{CCproblem}. Phenomenologically we face the coincidence problem \cite{conicidence}  and observationally there are several cosmological tensions related to the cosmological constant, like the Hubble tension \cite{hubbletension}, $\sigma_8$ tension \cite{sigmatension}, etc.

All these concerns for the cosmological constant representation of the dark energy, are enough to search for a dynamical alternative of dark energy \cite{odint}. This can be done in three major categories. The first one deals with generalizing the geometry part of the action. This can be done through different ways, leading to $f(R)$ gravities \cite{fR}, higher dimensional models \cite{higher}, massive gravities \cite{massive}, Weyl/Cartan/Finsler geometries \cite{WCF}, etc. In all the above models, the base assumption is that the geometry is not connected to gravity as simple as what we have in standard general relativity (GR). However, the main concern in these type of theories is the presence of ghost instabilities which can restrict the final form of the action \cite{ghost}.

The second category deals with modification of the matter sector. Here, we mainly assume that the negative pressure should be originated from some self-interaction of the matter fields. This leads to the scalar-vector theories of gravity \cite{scalarvector}, non-standard matter Lagrangians such as
$f(T)$, $f(T_{\mu\nu}T^{\mu\nu})$ \cite{modifiedmatter} and also derivatives of the matter Lagrangian \cite{derivativematter}. In the latter case, usually, the matter Lagrangian and its higher variations appear in the equation of motion which needs more attention to correctly define the matter energy-momentum tensor \cite{secondderivative}.

The third possibility is to modify the way matter and geometry interact. In this case, one considers non-minimal couplings between matter and geometry, leading to $f(R,T)$, $f(R_{\mu\nu}T^{\mu\nu})$, $f(R,L_m)$, etc. \cite{fRT}. The main concern here is a recent debate claiming that these types of models are not independent from scalar-vector-tensor gravity theories \cite{debate}. However, this debate has been fully answered in \cite{answer} and the main point is that in this third category, we consider the interaction of geometry and the baryonic matter sources from thermodynamics points of view making the resulting model independent from the scalar-vector-tensor gravity theories. In these types of theories, since matter couples non-minimally to geometry, generally the conservation of the energy-momentum tensor does not hold anymore. The consequence of the non-conservation is that geometry can be transformed to the baryonic matter field which will change the matter abundance of the late time universe. This could then be used to choose better model from observations \cite{mattercreation}.

One of the earliest attempts to modify general relativity, was done by Palatini \cite{palatini}, where the affine connection is taken to be an independent field. This implies that distances and parallel transports treated differently in general relativity. This basic idea was immediately proven to be identical to the standard metric-based GR and as a result served as a new method of variation in general relativity. However, these two variational methods no longer equivalent when one considers modified gravities from the first category. In fact, the Palatini-$f(R)$ gravity behaves very differently from metric-$f(R)$ model \cite{palatinifR}. So, one can choose the variational method to modify the way gravity interacts in the model. With this in mind, a new idea has recently attract some attentions, where these two variational methods are use together in a single theory \cite{hybrid}. The  resulting hybrid metric-Palatini model has a unique property that although we have a standard GR formulation, the independent connection still plays a significant role. In this theory, the metric that defines geometry is conformally related to the metric that defines the distances. As a result, one can write the theory as a new kind of scalar-tensor theory, mimicking the Brans-Dicke (BD) theory \cite{BD}. Cosmological and gravitational consequences of this theory is widely explored in the literature \cite{coshybrid}. 

In this paper, we will modify the way the matter and geometry interact with each other in the context of hybrid metric-Palatini model. This model then lies in the third category discussed above and as a result the resulting theory lacks conservation of the energy-momentum tensor. However, we will see that the conservation can be restored by assuming a specific matter Lagrangian on top of the FRW unvierse. Also, like the original hybrid metric-Palatini theory, the present model can also be written in scalar-tensor representation. Here, we have two interacting scalar fields, one is dynamical and interact BD-like to geometry and the other is non-dynamical and interact with the matter Lagrangian. This could be seen as a bi-scalar-tensor theory of gravity and the matter coupling could be seen as a chameleon field \cite{chameleon}. 

The paper is organized as follows. In the next section we will introduce the model and obtain the necessary field equations, including the conservation equation of the matter field. In Section \ref{sec3}, we construct the cosmological setup of the model and specify the matter–geometry interaction. In the following section, we perform the statistical analysis and examine the cosmological implications of the model. Section \ref{sec5} is devoted to the conclusions and final remarks.

\section{The model}\label{secII}
Let us consider an action functional of the form
\begin{align}\label{act}
	S = \kappa^{2}\int \mathrm{d}^{4} x \sqrt{-g} \Big[R + f(\mathcal{R}, L_{m})\Big]  + S_{m},
\end{align}
where $ R $ is the Ricci scalar constructed from the metric tensor, $ \mathcal{R}$ is the Palatini Ricci scalar constructed from  an independent affine connection $ \tensor{\hat{\Gamma}}{^\alpha_\mu_\nu} $ and $ S_{m} $ is the matter action defined as
\begin{align}
	S_m = \int\mathrm{d}^{4} x \sqrt{-g}L_m,
\end{align}
where $ L_{m} $ is the matter Lagrangian density constructed from matter fields and the metric tensor. Here, $f$ is an arbitrary function of the Palatini curvature and the matter Lagrangian, controlling the possible coupling between matter and geometry. It should also be noted that the matter sector is independent of the affine connection.

As was discussed before, the form of action \eqref{act} indicates a hybrid theory combining the metric and Palatini approaches. A non-minimal coupling between matter and geometry has a long history and its cosmological implications are vastly investigated. In this paper, we will consider the coupling between matter and geometry in the Palatini side of the model. Besides simplicity of the model compared to the standard metric-matter coupling theories, we will see that this choice will also affect the matter conservation equation. It should also be noted that non-minimal matter coupling in Palatini models have also been investigated in \cite{matterPalatini}.

It is customary in the Palatini and also hybrid metric-Palatini theories to write the action in a scalar-tensor form \cite{hybrid}. In order to do that, let us introduce two auxiliary fields $A$ and $B$ as 
\begin{align}\label{act2}
		S = \kappa^{2}& \int \mathrm{d}^{4} x \sqrt{-g} \big[  R + f(A,B)\nonumber\\ & + f_{A}(\mathcal{R} - A)  + f_{B}(L_{m} - B) \big] + S_{m},
\end{align}
where $ f_{X} $ represents derivative with respect to $X$. One can easily check by obtaining the equations of motion of $A$, $B$ that the action \eqref{act2} is equivalent to \eqref{act}. 

By defining the scalar fields 
\begin{align}\label{def1}
	 \phi\equiv f_{A} ,\quad  \psi\equiv f_{B}, 
\end{align}
and
\begin{align}\label{poten}
	 V(\phi, \psi) \equiv - f(A,B)+ \phi A + \psi B, 
\end{align}
the action \eqref{act2}  can be written in the form 
\begin{align}\label{act3}
	S = \int \mathrm{d}^{4} x \sqrt{-g} \Big[ \kappa^{2}  (R + \phi \mathcal{R}-V) + (1+\kappa^2\psi)L_{m} \Big].
\end{align}
This is an action for a bi-scalar-tensor theory of gravity. The non-minimal coupling of the scalar field $\phi$ is a well-known behavior in Palatini models. However, the scalar field $\psi$ introduces a new coupling between matter and geometry which is the main concern of the present paper.

In order to solve the theory for the Palatini part, let us take a variation of the action \eqref{act} with respect to the metric tensor and the independent connection as 
\begin{align}\label{eqmetr1}
	G_{\mu \nu} + \phi \mathcal{R}_{\mu \nu} + \frac12 g_{\mu \nu} \psi L_{m} - \frac12 g_{\mu \nu} f= \frac{1 + \kappa^{2} \psi}{2 \kappa^{2}} T_{\mu \nu}, 
\end{align}
and
\begin{align}\label{eqcon1}
	 \hat\nabla_{\alpha} (\sqrt{-g} \phi g^{\mu \nu}) = 0,
\end{align}
where we have used the definitions \eqref{def1}. Here, $\hat\nabla_\alpha$ represents covariant derivative with the independent connection $ \tensor{\hat{\Gamma}}{^\alpha_\mu_\nu} $.

Equation \eqref{eqcon1} suggests that the covariant derivative $\hat\nabla$ is compatible with a new metric defined as
\begin{align}\label{eqmetr2}
\sqrt{-h}h_{\mu \nu} =\sqrt{-g} \phi g_{\mu \nu}.
\end{align}
Using the $h$-metric, one can obtain the independent connection  $ \tensor{\hat{\Gamma}}{^\alpha_\mu_\nu} $ as
\begin{align}
	 \tensor{\hat{\Gamma}}{^\alpha_\mu_\nu}=\frac12h^{\alpha\beta}(\partial_\mu h_{\alpha\nu}+\partial_\nu h_{\mu\alpha}-\partial_\beta h_{\mu\nu}),
\end{align}
from which the Palatini Ricci scalar $\mathcal{R}$ could be constructed. Since the $h$-metric is conformally related to the $g$-metric, one can obtain the Palatini Ricci tensor in terms of the metric Ricci tensor as
\begin{align}\label{RvsR}
	 &\mathcal{R}_{\mu \nu} = R_{\mu \nu} + \frac{3}{2 \phi^{2}}  \partial_{\mu} \phi \partial_{\nu} \phi - \frac{1}{\phi} (\nabla_{\mu} \nabla_{\nu} \phi + \frac12 g_{\mu \nu} \Box \phi ).
\end{align}
Now, by substituting \eqref{RvsR} back into the action \eqref{act3}, one obtains a bi-scalar-tensor theory of gravity as
\begin{align}\label{act4}
		S &= \kappa^{2}\int \mathrm{d}^{4} x \sqrt{-g} \Big[  (1 + \phi) R  \notag \\ &+ \frac{3}{2 \phi} g^{\mu \nu} \partial_{\nu} \phi \partial_{\mu} \phi - V(\phi, \psi) + (\kappa^{-2}+ \psi) L_{m} \Big].
\end{align}
The above action, shows a dynamical scalar field $\phi$ non-minimally coupled to geometry and has the standard Brans-Dicke kinetic term with $\omega_{BD}=-3/2$. The scalar field $\psi$ on the other hand, lack the kinetic term, non-dynamically determines the relation between the potential $V$ and the matter sector. As a different viewpoint however, one can consider a theory with a varying gravitational constant $\kappa^{-2}_v = \kappa^{-2}+\psi$, acting as an auxiliary field. This resembles the Chameleon mechanism since the potential is implicitly matter field dependent \cite{chameleon}.

Equations of motion of the theory can be obtained by variation of the action \eqref{act4} with respect to the metric tensor $g_{\mu\nu}$, the scalar field $ \phi $ and the auxiliary field $ \psi $, with the result
\begin{align}\label{metreq}
 (1 &+ \phi) G_{\mu \nu} - \frac{3}{4 \phi} g_{\mu \nu} \nabla_{\alpha} \phi \nabla^{\alpha} \phi + \frac{3}{2 \phi} \nabla_{\mu} \phi \nabla_{\nu} \phi \notag \\ & + (g_{\mu \nu} \Box- \nabla_{\mu} \nabla_{\nu}) \phi + \frac12 g_{\mu \nu} V = \frac{1 + \kappa^{2} \psi}{2 \kappa^{2}} T_{\mu \nu},
\end{align}
\begin{align}\label{phieq}
	 R - \frac{3}{\phi} \Box \phi + \frac{3}{2 \phi^{2}} \nabla_{\alpha} \phi \nabla^{\alpha} \phi - V_{\phi} = 0,
\end{align}
and
\begin{align}\label{psieq}
		& L_{m} - V_{\psi} = 0.
\end{align}
As one can see from the above equations, since the matter sector is non-minimally coupled to the geometric field $\psi$, the conservation of the energy-momentum tensor does not hold anymore in this model. This can be easily proved by the Noether theorem under diffeomorphism invariance \cite{diffeomorphisminvariance}, or by explicitly taking covariant divergence of the metric field equation \eqref{metreq} with the result
\begin{align}\label{cons1}
	\nabla^{\mu} T_{\mu \nu} = \frac{\kappa^{2}}{1 + \kappa^{2} \psi} \big( T_{\nu \alpha} - g_{\nu\alpha} L_{m}\big) \nabla^{\alpha} \psi.
\end{align}  
It is evident that the non conservative term in the right hand side of equation \eqref{cons1} depends on the auxiliary field $\psi$. As a result the conservation of the energy-momentum tensor is retained in the case of vanishing $\psi$. There is however another special case where the conservation of the energy-momentum tensor holds. We will return to this issue in the next section.
\section{Cosmological implications}\label{sec3}
Let us assume that the universe can be described by an isotropic and homogeneous FRW metric of the form
\begin{align}\label{eq:12}
	\mathrm{d} s^2 = - \mathrm{d} t^2 + a^2(t) \big[ \mathrm{d} x^{2} + \mathrm{d} y^{2} + \mathrm{d} z^{2} \big],
\end{align}
where $ a(t) $ is the scale factor. Since the universe is homogeneous and isotropic, the matter content of the universe can safely be assumed to be a perfect fluid with Lagrangian density 
\begin{align}
L_m=-\rho,\label{lag1}
\end{align}
and energy-momentum tensor
\begin{align}\label{eq:13}
	T_{\mu\nu} =(\rho+p)u_\mu u_\nu+p g_{\mu\nu},
\end{align}
where $ \rho $ and $p $ are the energy density and thermodynamic pressure and $u^\mu$ is the fluid 4-velocity vector.

With the above assumptions, one can obtain from the field equation \eqref{metreq}, the modified Friedmann and Raychaudhuri equations as
\begin{align}
	& 3 H^{2} = \frac{1}{1 + \phi} \bigg[ \frac{1 + \kappa^{2} \psi}{\kappa^{2}} \rho + \frac{V(\phi, \psi)}{2} - 3 H \dot{\phi} - \frac34 \frac{\dot{\phi}^{2}}{\phi} \bigg], \label{eq:14} \\ &2 \dot{H} = \frac{1}{1 + \phi} \bigg[H \dot{\phi} + \frac32 \frac{\dot{\phi}^{2}}{\phi}
		- \ddot{\phi}- \frac{1 + \kappa^{2} \psi}{2 \kappa^{2}} (p + \rho)  \bigg], \label{eq:15}
\end{align}
respectively. The equations of motion of the scalar fields $\phi$ and $\psi$ could also be obtained as
\begin{align}
	& H^{2} = -2 \dot{H} - \frac34 \frac{\dot{\phi}}{\phi} + \frac18 \bigg( \frac{\dot{\phi}}{\phi} \bigg)^{2} - 4 \frac{\ddot{\phi}}{\phi} + \frac{1}{12}V_{\phi}, \label{eq:16} \\ & \rho + V_{\psi} = 0, \label{eq:17}
\end{align}
where we have used equations \eqref{phieq} and \eqref{psieq}. Equation \eqref{cons1} is also simplified to
\begin{align}
	\dot{\rho} +3 H (\rho+P )= 0, \label{eq:18}
\end{align}
indicating that the energy momentum tensor is conserved in this case. It should be noted that this conservation is a direct consequence of the FRW symmetry and the choice of Lagrangian density \eqref{lag1}. As is well-known in general relativity, the energy-momentum tensor of the perfect fluid can be obtained from two independent Lagrangians, e.g. $L_m=-\rho$ and $L_m=p$. However, these two choices result in different equations in modified gravity models with non-minimal matter-geometry couplings since the Lagrangian density appears explicitly in the field equations. For the present model, the right hand side of equation \eqref{cons1} can be simplified to
\begin{align}
	\textrm{RHS} = - \frac{\kappa^{2} \dot{\psi}}{1 + \kappa^{2} \psi} (L_{m} + \rho).
\end{align} 
One can see that with the choice \eqref{lag1}, the RHS of equation \eqref{cons1} vanishes, resulting in the conservation of the energy-momentum tensor. It should also be noted that this property is only true in the background cosmology and the model shows its non-conservative nature in the perturbative level, even with $L_m=-\rho$.

\subsection{Case I: The linear potential}\label{subsecA}
In order to solve the field equations \eqref{eq:15}-\eqref{eq:18}, one should first specify the form of the function $f(\mathcal{R},L_m)$. Let us first consider a minimal case with linear combination of the form
\begin{align}
	f(\mathcal{R},L_m) = \alpha \mathcal{R} + \beta L_m,
\end{align}
where $\alpha$ and $\beta$ are arbitrary constants.
In this case, one can easily prove from \eqref{poten} that the potential $V$ vanishes. From \eqref{eq:17} one can deduce that the model could not accept matter fields. As a result the Friedmann and $\phi$ equations will be simplified to
\begin{align}
	& 3 H^{2} = - \frac{1}{1 +‌\phi} \bigg[ \frac{\dot{\phi}}{4 \phi} (4 H \phi + \dot{\phi})  \bigg], \label{eq:20} \\
	& H^{2} = - 2 \dot{H} - \frac34 \frac{\dot{\phi}}{\phi} + \frac18 \bigg( \frac{\dot{\phi}}{\phi} \bigg)^{2} - 4 \frac{\ddot{\phi}}{\phi}. \label{eq:21}
\end{align}
It should be noted that the scalar-field $ \psi $ in this case is redundant since its is always coupled to the matter field which is vanishing here. 
In fact, the resulting equations \eqref{eq:20}, \eqref{eq:21} are equivalent to the hybrid metric-Palatini theory with vanishing matter field and potential \cite{hybrid}. 

By solving the equations \eqref{eq:20}, and \eqref{eq:21}, one can obtain the Hubble parameter and the scalar field $\phi$ as
\begin{align}
	 H(t) & = - \frac{1}{2( t + c_{2})}, \label{eq:22} \\ 
	 \phi(t) & = \frac{c_1}{\sqrt{2 (t + c_{2})}} - 1, \label{eq:23}
\end{align} 
where $ c_{1}>0 $ and $ c_{2} $ are the constant. Equation \eqref{eq:22} describe a universe that is expanding with a radiation-dominated behavior. Since we do not have matter sources in the model, this behavior comes from the scalar field $\phi$. In summary the resulting model is decelerating and can not be accepted for the late-time behavior of the universe.

\subsection{Case II: The multiplicative potential}\label{subsecB}
Let us consider a multiplicative interaction of the form
\begin{align}
	f = \alpha\mathcal{R}-\beta\mathcal{R}|L_m|^n,
\end{align}
where $\alpha$, $\beta$ and $n$ are some arbitrary constants
In this case, the potetial $ V $ could be obtained from equation \eqref{poten} as
\begin{align}
	V(\phi, \psi) = - \psi \left(\frac{ \alpha - \phi }  {\beta} \right)^{\frac1n}.
\end{align}
Substituting back to the field equations \eqref{eq:14}-\eqref{eq:17}, one obtains
\begin{align}
	3 H^{2} = \frac{1}{1 + \phi} \bigg[ \frac{1 + \kappa^{2} \psi}{\kappa^{2}} \rho  - \frac{\psi}{2} \bigg( \frac{\alpha - \phi}{\beta} \bigg)^{\frac{1}{n}} - 3 H \dot{\phi} - \frac34 \frac{\dot{\phi}^{2}}{\phi} \bigg], \label{eq:24} 
\end{align}
\begin{align}
	H^{2} = -2 \dot{H} - \frac34 \frac{\dot{\phi}}{\phi} + \frac18 \bigg( \frac{\dot{\phi}}{\phi} \bigg)^{2} - 4 \frac{\ddot{\phi}}{\phi} + \frac{\psi}{12 \beta n} \bigg( \frac{\alpha - \phi}{\beta} \bigg)^{\frac{1 - n}{n}}, \label{eq:25}
\end{align}
and
\begin{align}
\bigg(\frac{\alpha - \phi}{\beta}\bigg)^{\frac{1}{n}} = \rho. \label{eq:26}
\end{align}
From equation \eqref{eq:25} and \eqref{eq:26}, one could obtain the scalar fields $\phi$ and $\psi$ as
\begin{align}
	 \psi  &=3 \beta n \rho^{n - 1}  \bigg(4 H^{2}  + 2 \dot{H}  + \frac{3 \beta n (1 + w) \rho^{n} }{2\phi^{2}} \notag \\ & \times \Big[ 3 H^{2}  \Big(  \beta (n (1 + w)-2) \rho^{2} \nonumber\\&\qquad-2 \alpha (n (1 + w)-1) \Big) + 2\phi \dot{H}  \Big] \biggr), \label{eq:27}
\end{align} 
and
\begin{align}
	\phi =  \alpha - \beta \rho^{n}. \label{eq:28}
\end{align}
where we have used the conservation equation \eqref{eq:27} and $w$ is the equation of state parameter of the baryonic matter $p=w\rho$.

It is worth mentioning that any modified gravity model can be express as a dark energy model by rewriting the Friedmann and Raychaudhuri equations as
\begin{align}\label{Fr1}
	3H^2=\frac{1}{2\kappa^2} (\rho+\rho_{DE}),
\end{align}
and
\begin{align}\label{Fr2}
	2\dot{H}+3H^2=-\frac{1}{2\kappa^2}(p+p_{DE}),
\end{align}
where $\rho_{DE}$ and $p_{DE}$ are the dark energy (DE) energy density and pressure. 
In our case, the DE components can be obtained as
\begin{align}\label{rde}
	\frac{1}{2\kappa^2}\rho_{DE} =\frac{1}{1+\phi}\left[\frac{\kappa^{2} \psi-\phi}{\kappa^{2}} \rho + \frac{V}{2} - 3 H \dot{\phi} - \frac34 \frac{\dot{\phi}^{2}}{\phi}\right],
\end{align}
and
\begin{align}\label{pde}
	\frac{1}{2\kappa^2}p_{DE} =\frac{1}{1 + \phi} \left[\frac{\kappa^{2} \psi-\phi}{2 \kappa^{2}} p-\frac{V}{2}+2H \dot{\phi} - \frac34 \frac{\dot{\phi}^{2}}{\phi}
	+ \ddot{\phi}  \right].
\end{align}
Now, let us define the following set of dimensionless variables
\begin{align}\label{dimless1}
	\tau &=H_0t, \quad H=H_0 h,\quad \bar{\rho}=\f{1}{6\kappa^2H_0^2}\rho, \nonumber\\\beta &= \frac{\bar{\beta}}{6^{n} H^{2n}_{0} \kappa^{2n}},\quad \bar\psi = \kappa^2\psi,
\end{align}
and transform to redshift coordinate $z$ defined as
\begin{align}
1 + z = \frac{1}{a},
\end{align}
where $a$ is the scale factor. Assuming the universe is filled with dust with $w=0$, the conservation equation \eqref{cons1} leads to
\begin{align}
	\bar\rho(z)=\Omega_{m0}(1+z)^3,
\end{align}
where $\Omega_{m0}$ is the current value of the matter density abundance of the universe. With the above definitions, one can obtain the Hubble parameter as a function of redshift $z$ as
\begin{align}\label{hub1}
h(z) &= 2 (1 + z)^{\frac32} \left[\Omega_{m0}(\alpha - \Omega_{m1})\right]^\frac12\nonumber\\&\times\Big[4 (\alpha-\Omega_{m1})+ \big((2 - 3n)\Omega_{m1}-2\alpha\big)^2\Big]^{-\frac12},
\end{align}
where we have defined
\begin{align}
	\Omega_{m1}=\bar\beta (1 + z)^{3n} \Omega^{n}_{m0}.
\end{align}
In the case of vanishing $\beta$ or vanishing $n$, the above expression reduces to
$$h\propto (1+z)^\frac{3}{2},$$
which is identical to the case of linear potential we have discussed in the previous subsection. 

It should be noted that since $h(z=0)=1$ from the definition \eqref{dimless1}, one obtains the parameter $\alpha$ as
\begin{align}\label{alpha}
	\alpha = \frac12\bigg[\Omega_{m0}&-1-(3n-2)\bar\beta\Omega_{m0}^n\nonumber\\&-\sqrt{(1-\Omega_{m0})(1-\Omega_{m0}+6n\bar\beta\Omega_{m0}^n)}\bigg].
\end{align}
As a result, we have two model parameters $\bar\beta$ and $n$ which we will obtain from the cosmological data in the next section.

\section{Statistical analysis}\label{subsecC}
In order to obtain the evolution of the Hubble parameter from \eqref{hub1}, one should specify the optimal values for the cosmological and also model parameters $H_0$, $\Omega_{m0}$, $n$ and $\beta$. We will obtain the best fit values of these parameters by performing the Likelihood analysis using three different combinations of the following datasets.
\subsection{The Cosmic Chronometers}
The cosmic chronometer (CC) approach provides a direct and model-independent method for determining the Hubble parameter $H(z)$ by measuring the differential age evolution of passively evolving, massive early-type galaxies. In this method, the Hubble parameter is written as
\begin{equation}
	H(z) = -\frac{1}{1+z}\frac{dz}{dt}.
\end{equation}
The value of Hubble parameter $H(z)$, can then be inferred from the measurement of age difference between two nearby galaxies separated by a small redshift interval $\Delta z$. This method is independent of the cosmological model. In this paper we employ the 31 CC data points \cite{CCdata} which assumed to be independent. It should be noted that the covariance matrix between these data can also be found by assuming a specific stellar population synthesis (SPS) model for the underlying galaxy population \cite{popCC} which we will not consider in this paper since it partially breaks the model-independence of CC method.
The contribution of CC dataset to the total likelihood is
\begin{align}
	\chi^{2}_{\text{CC}} = \sum_i \left( \frac{H_{\text{obs},i} - H_{\text{th},i}}{\sigma_i} \right)^2,
\end{align}
where $i$ labels the data points, $H_{\text{obs},i}$ are the observational estimates of the Hubble parameter reconstructed from differential ages, $H_{\text{th},i}$ are the theoretical predictions of the model at corresponding redshifts, and $\sigma_i$ denotes the reported 1$\sigma$ uncertainties.

\subsection{The Pantheon$^+$}
The Pantheon$^+$ compilation \cite{PANdata} represents the most updated and homogeneous collection of Type Ia supernova (SN Ia) distance measurements, consists of about 1500 spectroscopically confirmed SNe~Ia spanning the redshift range $0.001 < z < 2.26$, combining observations from 18 different surveys. Pantheon$^+$ improves the Pantheon dataset by enhancing the photometric calibration and refining light-curve fitting.
Here we employ the Pantheon$^+$ dataset without the SH0ES Cepheid calibration \cite{SH0ES}, so the absolute magnitude $M$ is assumed to be a free parameter which we will inferred from the fitting. 
The Pantheon$^+$ measurements are not independent and the covariance matrix is provided in \cite{PANdata}. 
The contribution of the Pantheon$^+$ dataset to the total likelihood is
\begin{equation}
	\chi^2_{\mathrm{Pantheon}^+}
	=
	\left[ \vec{\mu}_{\mathrm{obs}} - \vec{\mu}_{\mathrm{th}} \right]^{T}
	C^{-1}
	\left[ \vec{\mu}_{\mathrm{obs}} - \vec{\mu}_{\mathrm{th}} \right],
\end{equation}
where $C$ is the covariance matrix of the Pantheon$^+$ data.

\subsection{The DESI(DR2) BAO}
We use the Baryon Acoustic Oscillation (BAO) measurements from the second data release of the Dark Energy Spectroscopic Instrument (DESI DR2) \cite{DESIDR2}. The dataset includes BAO observables extracted from several tracers of large-scale structure, namely the Bright Galaxy, the Luminous Red Galaxy and the Emission Line Galaxy samples and also quasars, covering the interval $0.2 \lesssim z \lesssim 2.4$. The DESI DR2 analysis reports BAO measurements in terms of distance ratios, typically of the form
\begin{equation}
	\frac{D_{\mathrm{M}}(z)}{r_{\mathrm{d}}}, \qquad
	\frac{D_{\mathrm{H}}(z)}{r_{\mathrm{d}}}, \qquad
	\frac{D_{\mathrm{V}}(z)}{r_{\mathrm{d}}},
\end{equation}
where $D_{\mathrm{M}}(z)$ is the comoving angular diameter distance, $D_{\mathrm{H}}(z)$ is the Hubble distance, and $D_{\mathrm{V}}(z)$ is the spherically averaged distance. The quantity $r_{\mathrm{d}}$ denotes the sound horizon at the drag epoch which we will assume to be a free parameter and infer from the fitting process. Defining the vector
\[
\vec{X}= 
\left(\frac{D_{\mathrm{M}}(z)}{r_{\mathrm{d}}},\; \frac{D_{\mathrm{H}}(z)}{r_{\mathrm{d}}},\; \frac{D_{\mathrm{V}}(z)}{r_{\mathrm{d}}} \right),
\]
the BAO contribution to the likelihood is then given by the $\chi^2$ function as
\begin{equation}
	\chi^{2}_{\mathrm{BAO}}
	=
	\left[ \vec{X}_{\mathrm{obs}} - \vec{X}_{\mathrm{th}} \right]^{T}
	C^{-1}
	\left[ \vec{X}_{\mathrm{obs}} - \vec{X}_{\mathrm{th}} \right],
\end{equation}
where $C$ is the covariance matrix.

In this paper, we will use three different combinations of the datasets, namely
\begin{itemize}
	\item CC+Pantheon$^+$,
	\item CC+BAO,
	\item CC+Pantheon$^+$+BAO,
\end{itemize}
The likelihood function can then be defined as
\begin{align}
	L=L_0e^{-\chi^2/2},
\end{align}
where $L_0$ is the normalization constant with corresponding loos function for each case. 

By maximizing the likelihood function, the best fit values of the parameters $H_0$, $\Omega_{m0}$, $n$, $\beta$, $\mathcal{M}$ and and $r_d$ at $1\sigma$ confidence level is summarized in table \ref{bestfit}.
We have also reported the fitted values of the derived parameter $\alpha$ defined in \eqref{alpha} and also the reduced $\chi^2_{red}$ defined as
$$\chi^2_{red}=\frac{\chi^2}{dof},$$
where $dof$ is the number of degrees of freedom. From the table one can find that we have a good fit for datasets $CC+Pantheon^+$ and $CC+Pantheon^++BAO$. However, the $CC+BAO$ dataset is overfitted due to small number of data points.
\begin{table*}[t]
	\centering
	\renewcommand{\arraystretch}{1.3}
	\begin{tabular}{|c||c|c|c||c|c|c|}
		\hline
		\multirow{2}{*}{\textbf{Parameter}} 
		& \multicolumn{3}{c||}{\textbf{$\Lambda$CDM model}} 
		& \multicolumn{3}{c|}{\textbf{Non-minimal HMP model}} \\ \cline{2-7}
		& \textbf{CC+PNT} & \textbf{CC+BAO} & \textbf{CC+PNT+BAO} 
		& \textbf{CC+PNT} & \textbf{CC+BAO} & \textbf{CC+PNT+BAO} \\ \hline\hline
		\rule{0pt}{12pt} $\mathrm{H_{0}}$ 
		& $66.552_{-1.663}^{+1.701}$ & $69.083_{-1.577}^{+1.642}$ & $68.499_{-1.612}^{+1.607}$ 
		& $67.082_{-1.708}^{+1.699}$ & $69.715_{-2.187}^{+2.181}$ & $67.040_{-1.535}^{+1.602}$ \\ \hline
		\rule{0pt}{12pt} $\mathrm{\Omega_{m0}}$ 
		& $0.3570_{-0.0176}^{+0.0180}$ & $0.2981_{-0.0084}^{+0.0085}$ & $0.3103_{-0.0080}^{+0.0079}$ 
		& $0.2712_{-0.0484}^{+0.0496}$ & $0.2920_{-0.0134}^{+0.0125}$ & $0.3169_{-0.0154}^{+0.0146}$ \\ \hline
		\rule{0pt}{12pt} $\mathrm{\beta}$ 
		& \text{--} & \text{--} & \text{--} 
		& $1.211_{-0.423}^{+0.420}$ & $2.333_{-0.712}^{+0.825}$ & $1.157_{-0.194}^{+0.204}$ \\ \hline
		\rule{0pt}{12pt} $\mathrm{n}$ 
		& \text{--} & \text{--} & \text{--} 
		& $0.689_{-0.043}^{+0.041}$ & $0.697_{-0.020}^{+0.020}$ & $0.723_{-0.025}^{+0.022}$ \\ \hline
		\rule{0pt}{12pt} $\mathrm{\mathcal{M}}$ 
		& $-19.452_{-0.053}^{+0.053}$ & \text{--} & $-19.404_{-0.050}^{+0.050}$ 
		& $-19.426_{-0.054}^{+0.055}$ & \text{--} & $-19.431_{-0.048}^{+0.051}$ \\ \hline
		\rule{0pt}{12pt} $\mathrm{r_d}$ 
		& \text{--} & $146.997_{-3.418}^{+3.337}$ & $146.874_{-3.409}^{+3.353}$ 
		& \text{--} & $146.792_{-3.444}^{+3.493}$ & $146.821_{-3.342}^{+3.373}$ \\ \hline\hline
		\rule{0pt}{12pt} $\mathrm{\alpha}$ 
		& \text{--} & \text{--} & \text{--} 
		& $-1.0906_{-0.0981}^{+0.1192}$ 
		& $-1.3258_{-0.1607}^{+0.1866}$ 
		& $-1.0838_{-0.0616}^{+0.0674}$ \\ \hline
		\rule{0pt}{12pt} $\chi^2_{red}$ 
		& $1.044$ & $0.608$ & \text{1.046} 
		& $1.046$ 
		& $0.693$ 
		& $1.044$ \\ \hline
		
	\end{tabular}
	\caption{The parameter constraints of the $\Lambda$CDM and non-minimal HMP models for the CC+Pantheon$^+$, CC+BAO abd CC+Pantheon$^+$+BAO dataset combinations. We have also reported the fitted value of the derived parameter $\alpha$ and also the reduced $\chi^2_{red}$ for completeness.}
	\label{bestfit}
\end{table*}

\begin{figure*}
	\includegraphics[scale=0.4]{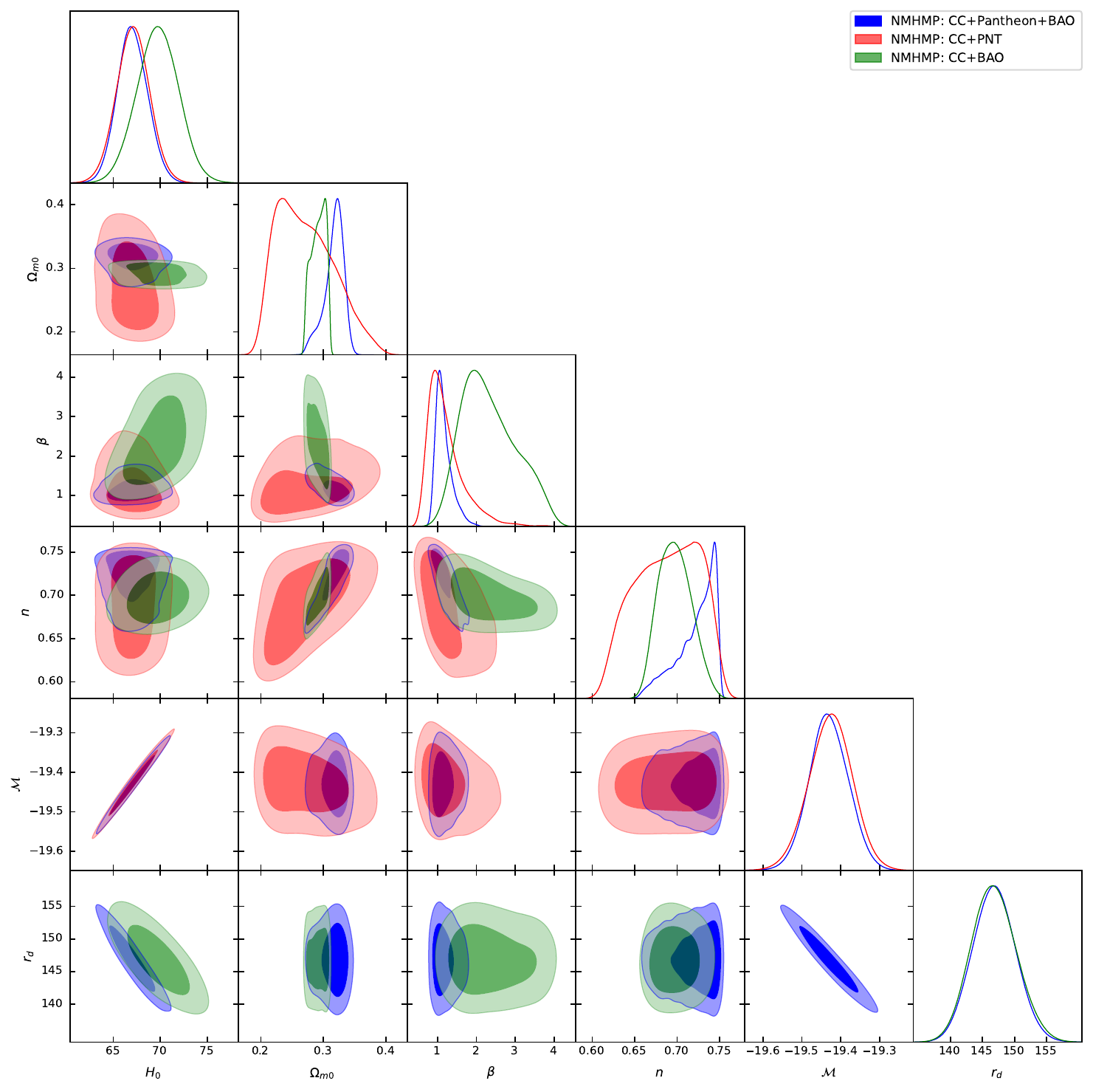}\includegraphics[scale=0.4]{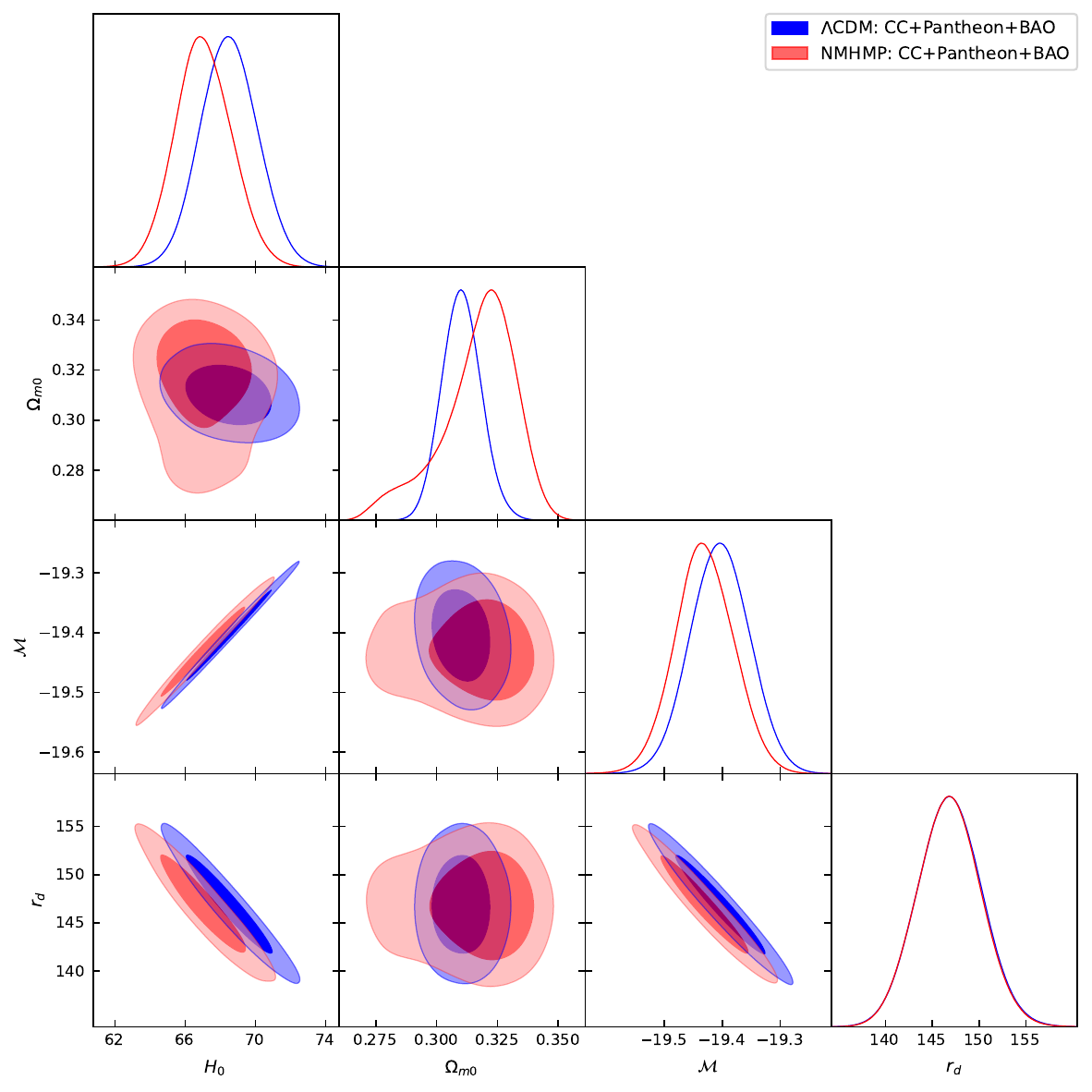}
	\caption{\label{cornerMG} The corner plot of the values of parameters with their $1\sigma$ and $2\sigma$ confidence levels for the NMHMP model (left) and together with $\Lambda$CDM model (right).}
\end{figure*}

\begin{figure}[h!]
	\includegraphics[scale=0.45]{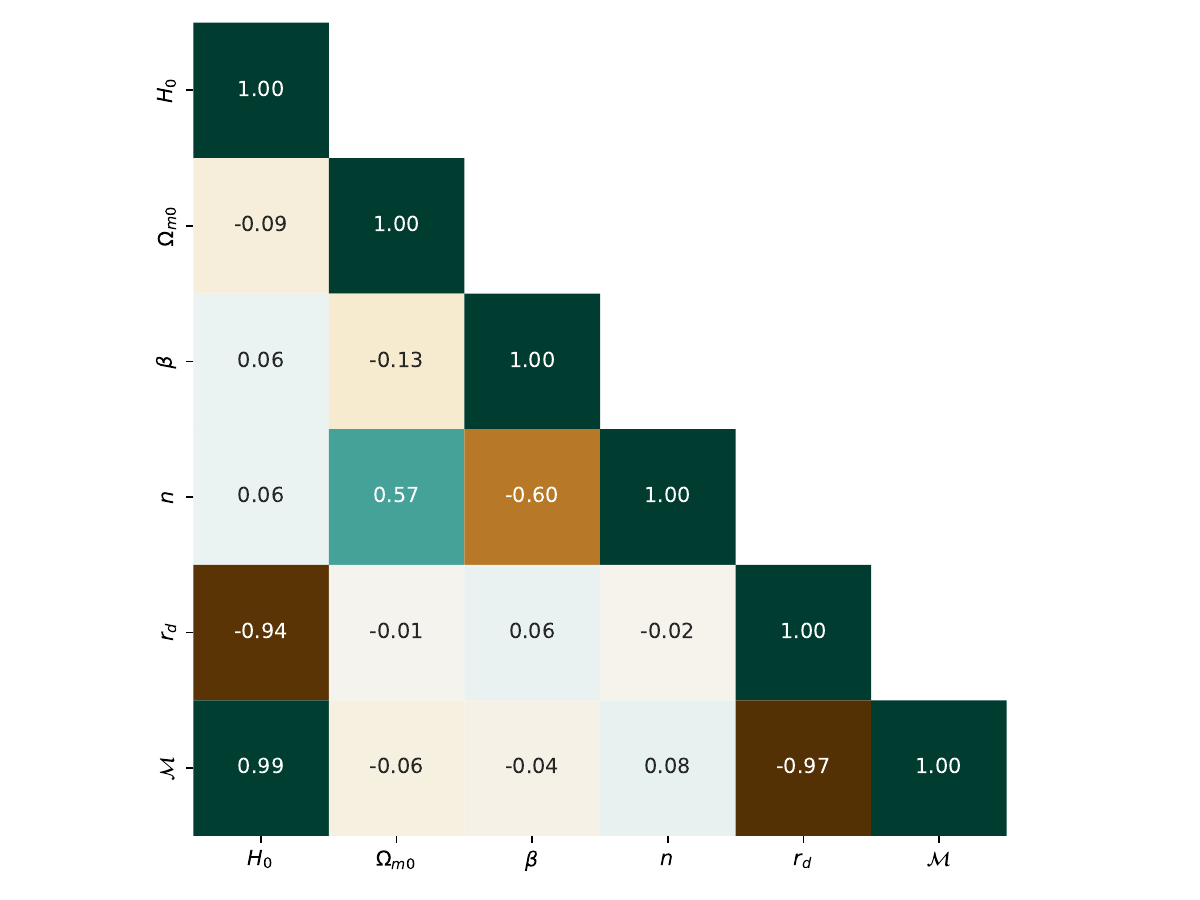}
	\caption{\label{corr} The Pearson correlation matrix of between the parameters of the NMHMP model. Moderate correlation of the model parameter $n$ with $\beta$ and $\Omega_{m0}$ can be seen from the figure.}
\end{figure}

The corner plot for the values of parameters with their $1\sigma$ and $2\sigma$ confidence levels for the non-minimal hybrid metric-Palatini (NMHMP) is shown in figure \eqref{cornerMG}. We have also reported the corner plot of the cosmological parameters for both NMHMP and $\Lambda$CDM models.
As one can see from the figures, the model parameters $n$ and $\beta$ are correlated. In figure \eqref{corr}, we have plotted the Pearson correlation matrix for the parameters $H_0$, $\Omega_{m0}$, $n$, $\beta$, $\mathcal{M}$ and $r_d$. The Pearson correlation matrix quantifies the degree of linear dependence between the parameters and defined as
\begin{align}
	r_{ij} = \frac{\textmd{cov}(x_i,x_j)}{\sigma_i\sigma_j},
\end{align}
where $r_{ij}$ denotes the $ij$ component of the matrix obtained from the parameter pair $(x_i,x_j)$. Here $\textmd{cov}$ is the covariance and $\sigma$ is the standard deviation.
One can see that the model parameter $n$ is moderately correlated with parameters $\beta$ and also with the matter abundance $\Omega_{m0}$. Increasing the value of the parameter $n$ leads to a decrease in the parameter $\beta$ and an increase in $\Omega_{m0}$. As a result, larger values of the exponent $n$ will result in more matter abundance in the present energy budget of the universe. It is in fact reasonable since increasing $n$ makes the non-minimal coupling stronger, thereby producing more matter. It should also be noted that there is also a strong correlation between the cosmological parameters $\mathcal{M}$ and $r_d$. Such a result is generic in cosmology, and the NMHMP model exhibits the same behavior.
\subsection{The cosmological results}
\begin{figure}[h!]
	\includegraphics[scale=0.47]{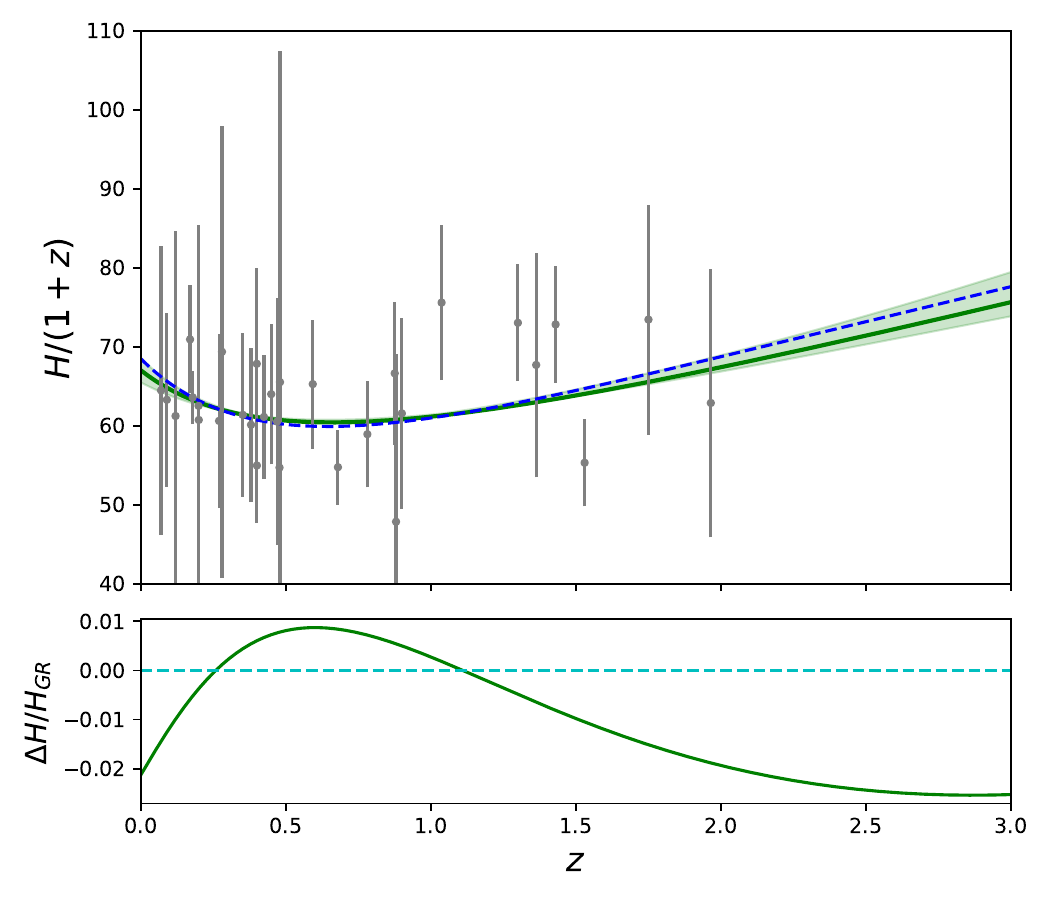}
	\caption{\label{fighubdif} The behavior of the rescaled Hubble parameter $H/(1+z)$ (top panel) and the difference between the two models (bottom panel) as a function of the redshift for the best fit values of the parameters as given by table \ref{bestfit}. The shaded area denotes the $1\sigma$ error. Dashed lines represent $\Lambda$CDM model. The error bars correspond to the observational data of the Cosmic Chronometers dataset.}
\end{figure}

Let us now consider the late-time cosmological evolution of the NMHMP  model using the results obtained in the previous subsection. In the following, we will only use the complete $CC+Pantheon^++BAO$ result. In figure \eqref{fighubdif} we have plotted the Hubble parameter as a function of redshift $z$ together with the relative difference of the Hubble parameters in NMHMP and $\Lambda$CDM models.
First of all, one can see that the NMHMP theory behaves very similar to the $\Lambda$CDM model. However, there are some minor differences that we are going to explore in details. The minimum of the Hubble parameter diagram represents the time of deceleration/acceleration transition.  It can be seen from the figure that this transition takes place at earlier times (larger redshifts) than the $\Lambda$CDM model. This means that the accelerated phase of the universe in NMHMP theory is a little older than its $\Lambda$CDM counterpart. Also, one can see that the NMHMP plot lies below the $\Lambda$CDM curve most of the time, getting above only in redshifts $z\in(0.5,1)$. Usually, phantom/quintessence dark energy models has a property that its Hubble plot lies below/above the $\Lambda$CDM curve. This suggest qualitatively that the NMHMP theory describes a dark energy model with a phantom-quintessence transition.  In the bottom part of the figure \eqref{fighubdif}, we have also plotted the difference between the Hubble parameters of NMHMP and $\Lambda$CDM models, showing that the difference is growing in the early times.

\begin{figure}[h]
	\includegraphics[scale=0.5]{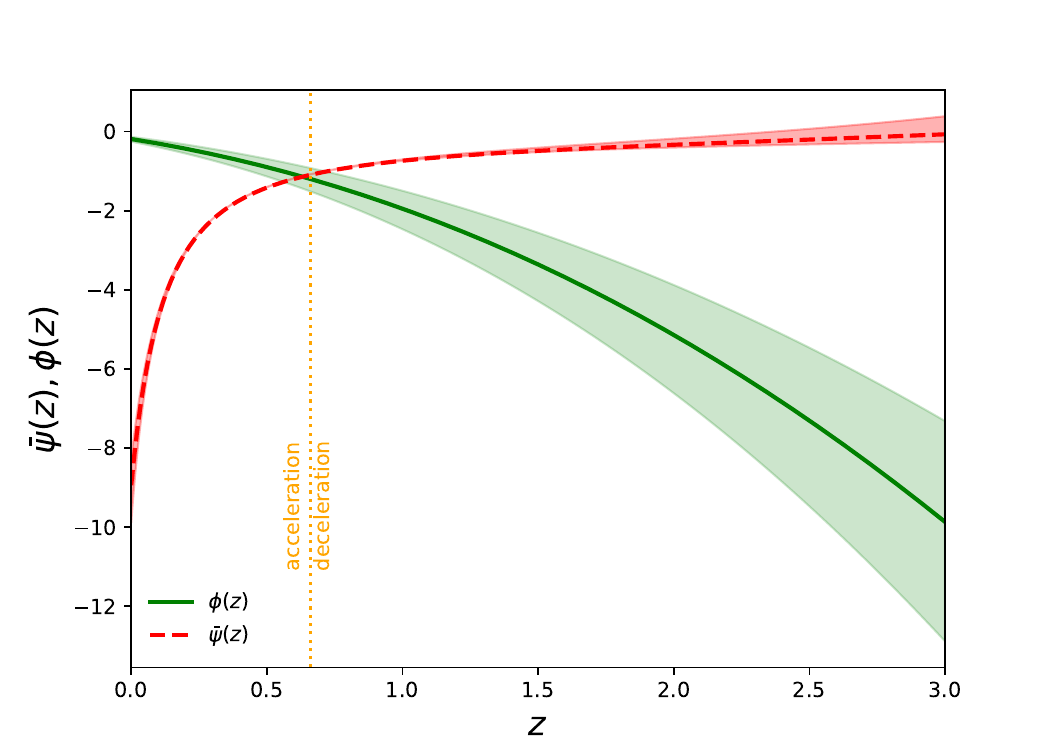}
	\caption{\label{figpsiphi} The behavior of scalar fields $\bar\psi$ (dashed line) and $\phi$ (solid line) as functions of the redshift for NMHMP model for the best fit values of the parameters as given by table \ref{bestfit}. The vertical dashed line represent acceleration to deceleration transition redshift.}
\end{figure}
In figure \eqref{figpsiphi}, we have plotted the behavior of the scalar fields $\phi$ and $\bar\psi$ as functions of the redshift $z$. As was mentioned after the action \eqref{act3}, the scalar field $\bar\psi$ controls the non-minimal matter-geometry coupling and the scalar field $\phi$ coupled to the Palatini scalar curvature. One can see from the figure that the scalar field $\bar\psi$ is nearly vanishing through time, starting to decrease at late times. This implies that the non-minimal matter-geometry coupling in the NMHMP model is only important at late times and responsible for the late time accelerated expansion of the universe. On the other hand, the scalar filed $\phi$ decreases with redshift, becoming nearly zero at the present time. As a result, the scalar field $\phi$ is less responsible for the acceleration of the late time universe.

\begin{figure}[h!]
	\includegraphics[scale=0.47]{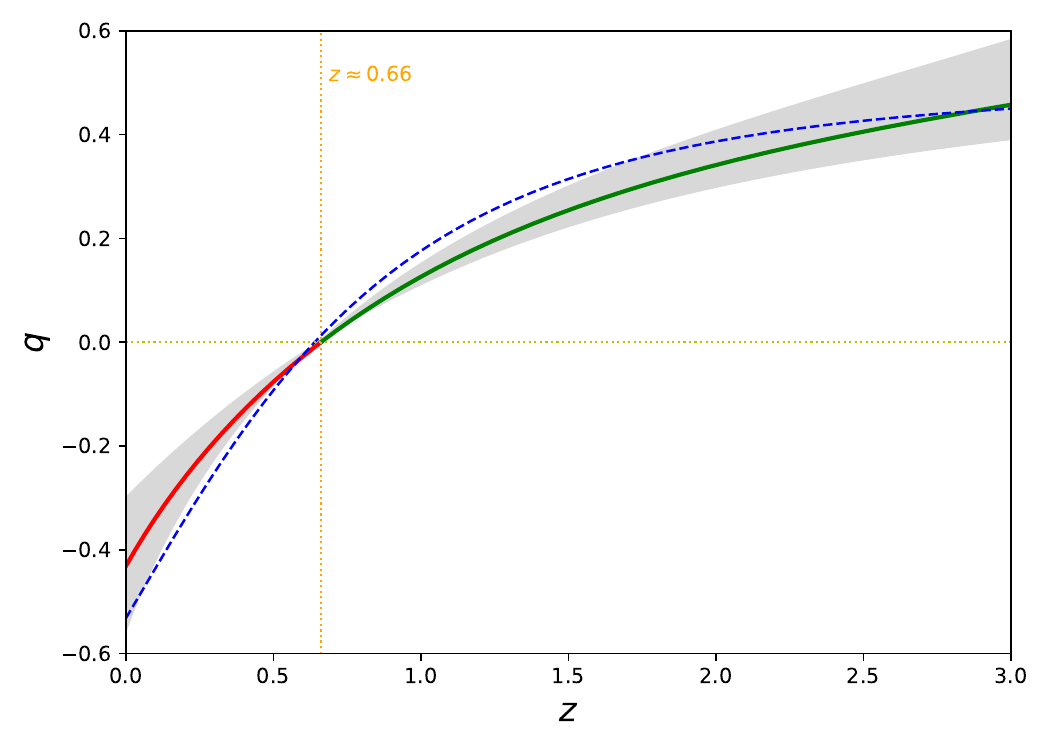}
	\caption{\label{figdec} The behavior of the deceleration parameter $q$as a function of the redshift $z$ for NMHMP model for the best fit values of the parameters as given by table \ref{bestfit}. The shaded area denotes the $1\sigma$ error. Dashed lines represent $\Lambda$CDM model. The vertical dashed line denotes the deceleration to acceleration phase transition redshift.}
\end{figure}

In order to dive deeper into the transitions discussed above, let us consider higher order derivatives of the Hubble parameter. In figure \eqref{figdec} we have plotted the deceleration parameter defined as
\begin{align}
	q = -1+(1+z)\frac{h^\prime}{h}.
\end{align}
\begin{figure*}
	\includegraphics[scale=0.5]{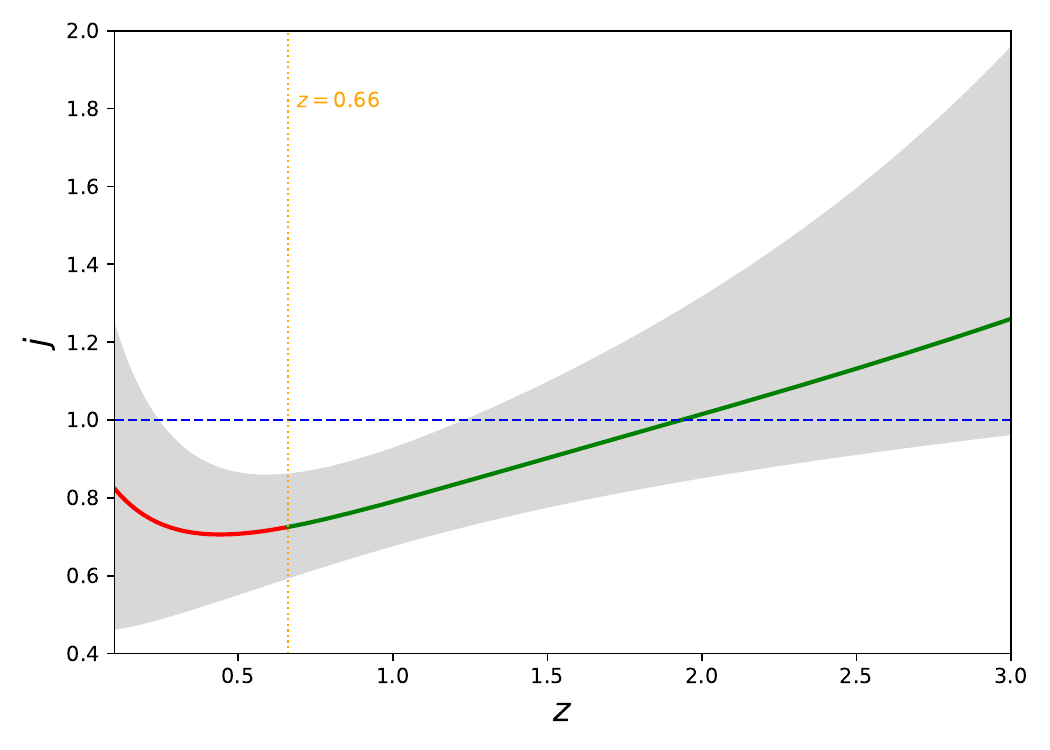}\includegraphics[scale=0.5]{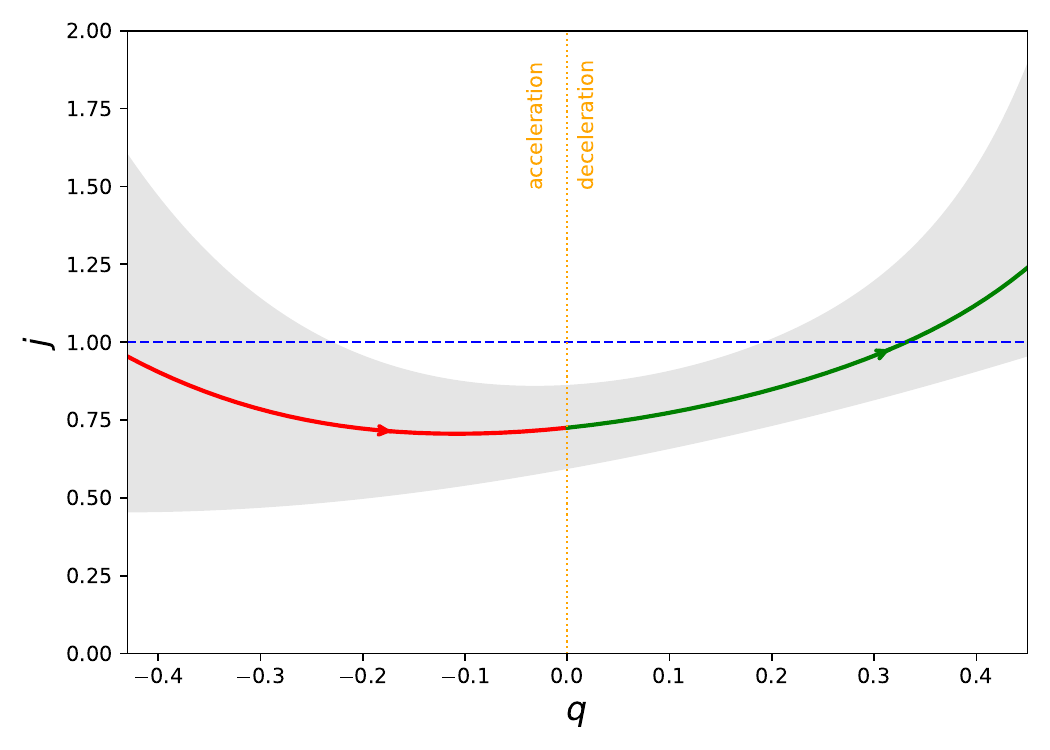}
	\caption{\label{figjerkjerk} The behavior of jerk as a function of redshift $z$ (left) and as a function of the deceleration parameter $q$ (right) for the NMHMP gravity model for the best fit values of the parameters as given by table \ref{bestfit}. The shaded area denotes the $1\sigma$ error. Dashed lines represent $\Lambda$CDM model.}
\end{figure*}
The dashed line represents the $\Lambda$CDM behavior. One can see from the figure that the deceleration to acceleration transition takes place at redshift $z\approx0.66$ and as we have noted is earlier than its $\Lambda$CDM counterpart which occurs at $z\approx0.64$. Also, it is evident that the acceleration of the universe in NMHMP theory is smaller than the $\Lambda$CDM model at present time. From the deceleration parameter, one can also infer another important information. As can be seen from the figure, we have $q>-1$ signaling the behavior of quintessence for the model. This suggests that the phantom era in the dark energy sector is not  strong enough to make the whole universe experience phantom-like acceleration. We will come back to this point in the following.  As a final point we mention that the NMHMP model predicts also lower deceleration in the matter dominated phase of the expansion. In summary, one can see that the non-minimal coupling in this model behaves slightly weaker than the $\Lambda$CDM model.

In figure \eqref{figjerkjerk} we have plotted the jerk parameter $j$ which can be written in redshift coordinates as
\begin{align}
	 j(z) = 2q^2 + q + (1 + z)\frac{dq}{dz},
\end{align}
as a function of redshift and also as a function of the deceleration parameter.

 The jerk parameter as a third derivative of the scale factor can determine the slope of the Hubble parameter curve. Larger values of the jerk parameter, represent higher slope and an increase in acceleration of the universe. The horizontal dashed line in figures \eqref{figjerkjerk} are the $\Lambda$CDM behavior and we have split the deceleration/acceleration regimes. One can see from the figure that the NMHMP model has larger jerk at earlier times than the $\Lambda$CDM model, becoming smaller at late times. Usually models with phantom/quintessence dark energy component has a bigger/smaller jerk parameter than their $\Lambda$CDM counterparts. As was also inferred from previous plots, we have a $\Lambda$CDM crossing in the NMHMP theory suggesting that the dark energy component of the model has a phantom-quintessence transition. It is also evident from these figures that the phantom behavior should occur at earlier times.

The true and exact dark energy (DE) behavior of the NMHMP model can be obtained by analyzing the DE energy density and pressure \eqref{rde}-\eqref{pde}.
\begin{figure*}
	\includegraphics[scale=0.53]{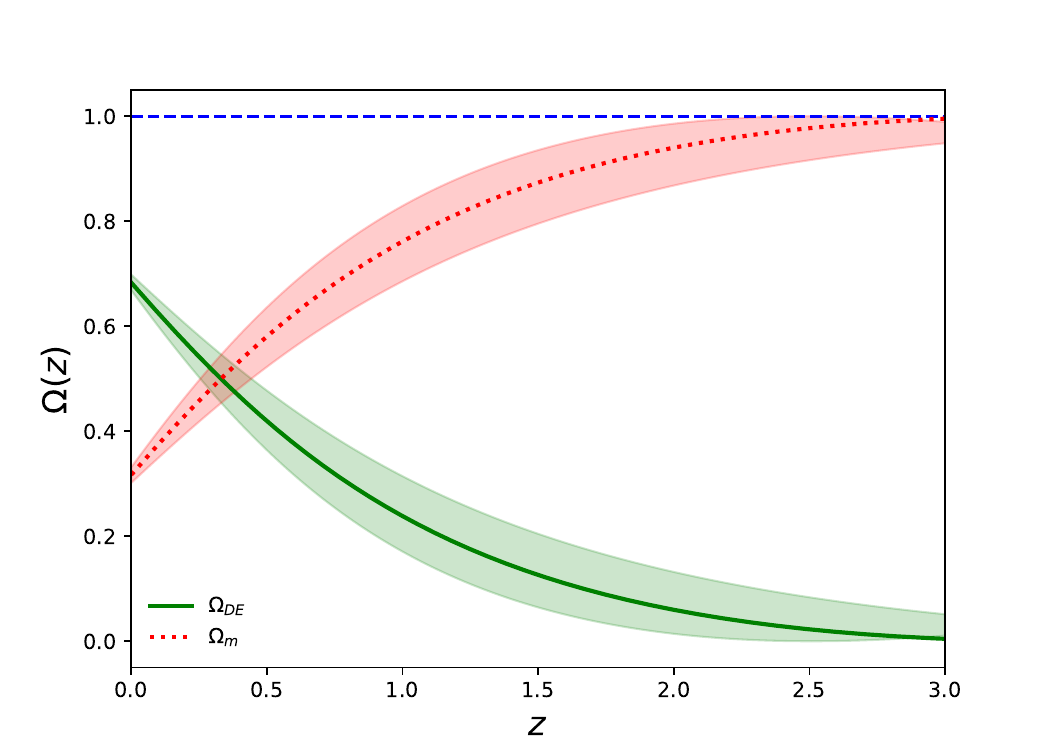}\includegraphics[scale=0.53]{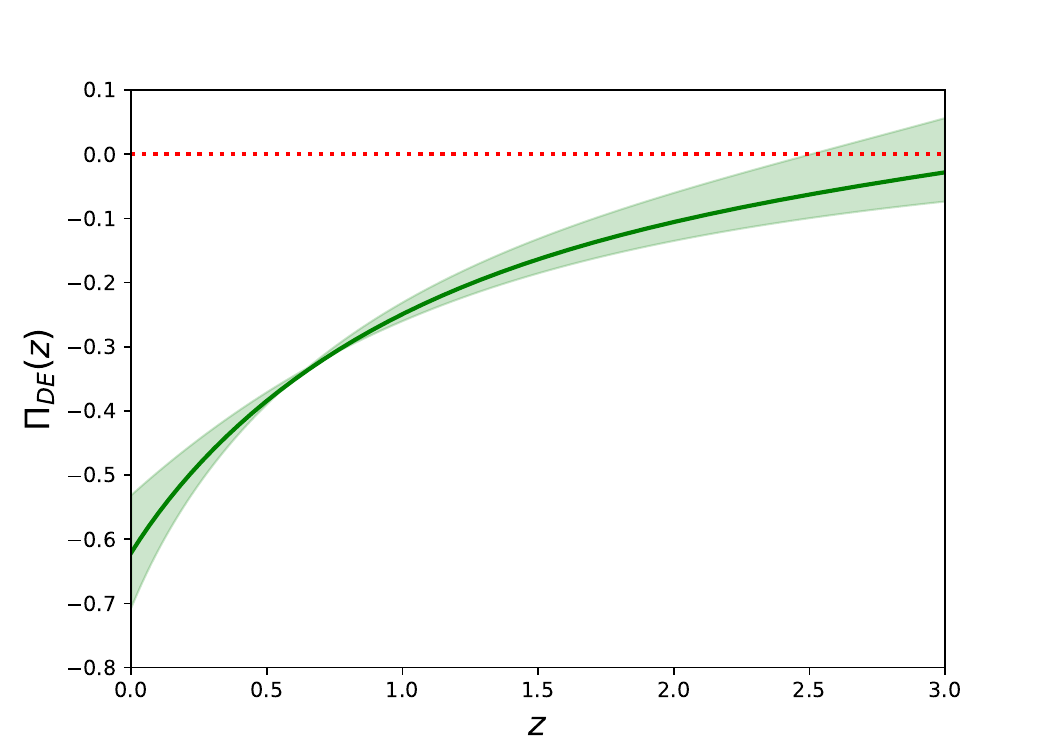}
	\caption{\label{figrhoandp} The behavior of the DE energy density (left panel) and the DE pressure (right panel) as a function of the redshift for the NMHMP model for the best fit values of the parameters as given by table \ref{bestfit}. The shaded area denotes the $1\sigma$ error. Dashed line represents the total value of the matter abundance and the dotted lines denote the baryonic matter contribution.}
\end{figure*}
In figure \eqref{figrhoandp} we have plotted the behavior of the energy density and pressure of the dark energy sector as a function of the redshift. In the energy density diagram, the increasing (dotted) line represents the behavior of the baryonic matter and the decreasing (solid) line represents the DE behavior, summed up to unity (the dashed line). One should note that the plotted curves are universal in all DE models with conservative matter sector; dark energy density increases as we reach the present time causing the accelerated expansion of the universe. The DE pressure on the other hand is a negative increasing function of the redshift, resembling the well-known behavior of cosmological constant with $p_{DE}\propto-\rho_{DE}$. 

\begin{figure*}
	\includegraphics[scale=0.5]{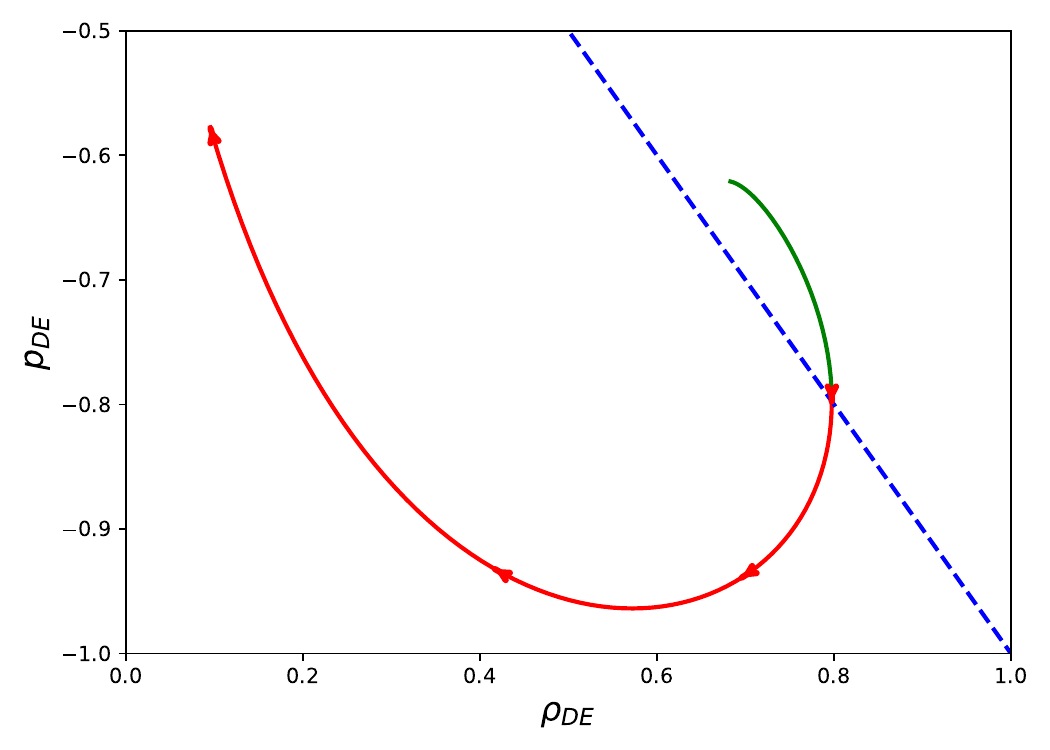}\includegraphics[scale=0.5]{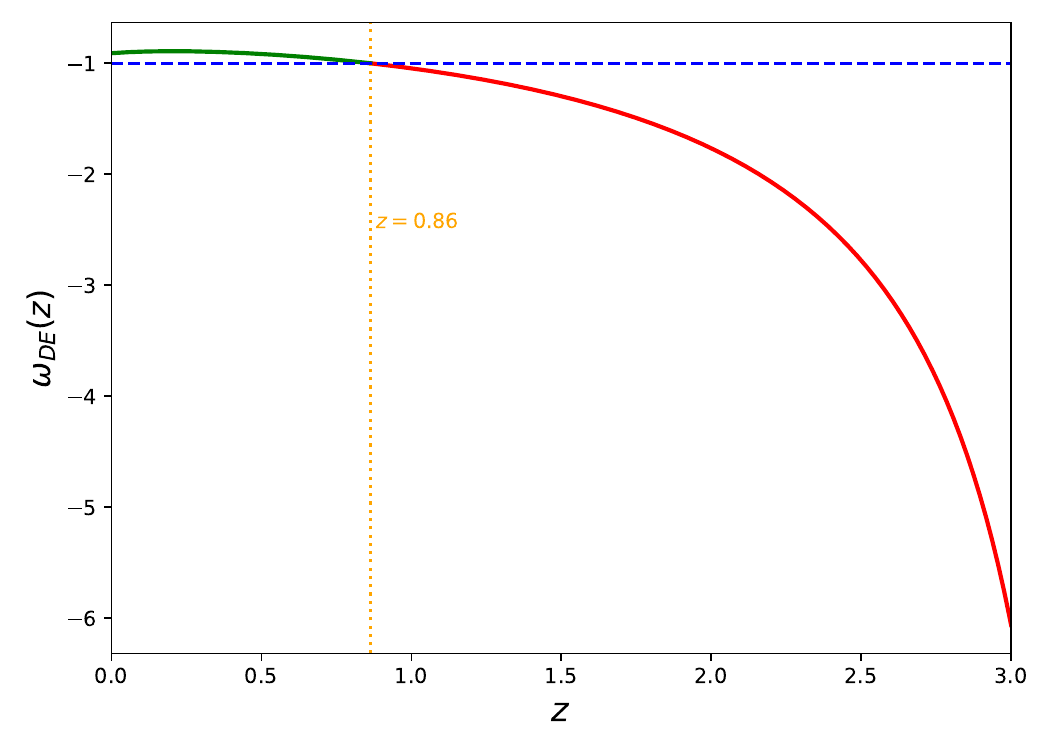}
	\caption{\label{figomegaDE} The evolution of DE pressure as a function of DE energy density (left panel) and of the DE equations of state parameter as a function of redshift (right panel) for the best fit values of the parameters as given by table \ref{bestfit}. The dashed lines represent the behavior of the $\Lambda$CDM model. Also, the quintessence-phantom transition redshift is denoted as a vertical line in the right plot.}
\end{figure*}

In order to understand the detailed behavior of DE sector, in figure \eqref{figomegaDE} we have plotted the evolution of DE pressure as a function of DE energy density and also the equation of state of the DE sector $\omega_{DE}$ defined as
\begin{align}
	\omega_{DE}=\frac{p_{DE}}{\rho_{DE}}.
\end{align}
It is evident from the figure that the DE sector in fact experience a quintessence to phantom transition at redshifts about $z\approx0.86$. As a result the DE sector is mostly phantom-like, changing to quintessence at late times. Also, one can see from these figures that the DE behaves very similar to the cosmological constant at late times; mostly in the quintessence regime.

\begin{figure}
	\includegraphics[scale=0.5]{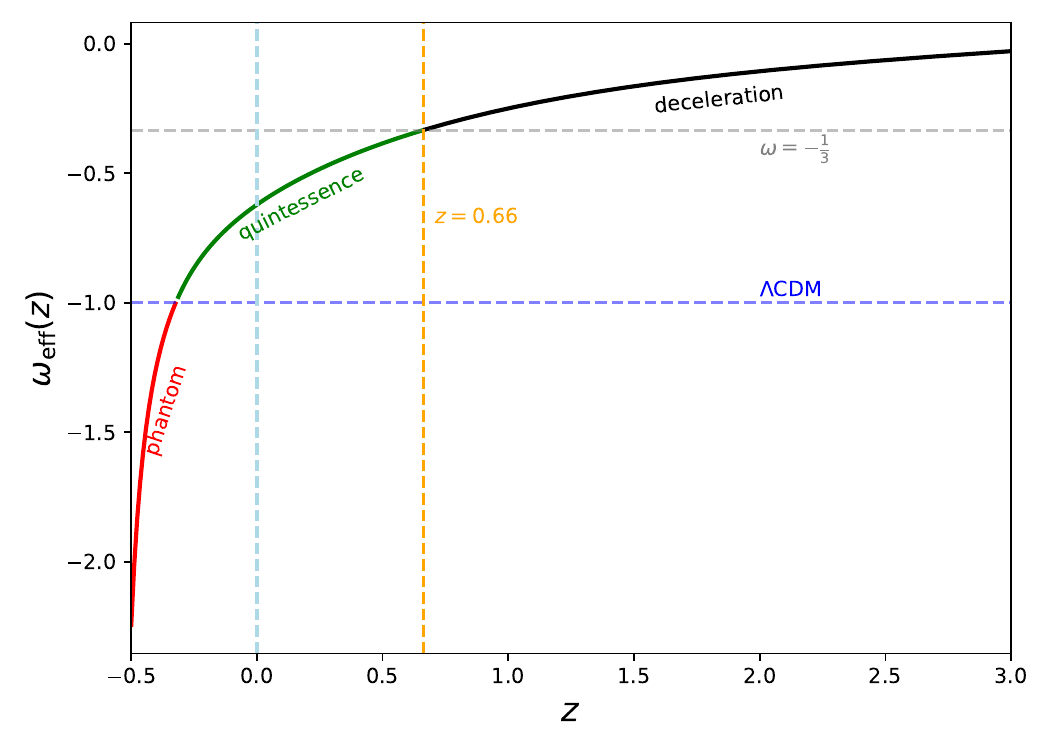}
	\caption{\label{figomegaeff} The behavior of the effective equation of state parameter as a function of the redshift $z$ for the NMHMP model for the best fit values of the parameters as given by table \ref{bestfit}.}
\end{figure}

Although the DE sector behaves phantom-like at late times, the effective behavior of the whole universe is governed by both DE and baryonic sectors. In figure \eqref{figomegaeff}, we have plotted the effective equation of state parameter, defined as
\begin{align}
	\omega_{eff} = \frac{p_{DE}}{\rho_m+\rho_{DE}},
\end{align}
as a function of redshift $z$. As one can see from the figure, the deceleration to acceleration transition takes place at redshifts around $z\approx0.66$ as we have obtained earlier from the deceleration parameter. Also, the universe accelerates quintessence-like at the present time. We have also plotted the behavior of the universe in the near future, with $z<0$, As can be inferred from the figure, the universe will accelerates phantom-like in the future.

A well-known diagnostic tool for analyzing the behavior of the DE sector is the statefinder analysis introduced in \cite{statefinder}. In this analysis we use the statefinder pair $\{j,\bar{s}\}$ where $j$ is the jerk parameter and $\bar s$ is a combination of the deceleration parameter $q$ as $j$ defined as
\begin{figure}
	\includegraphics[scale=0.5]{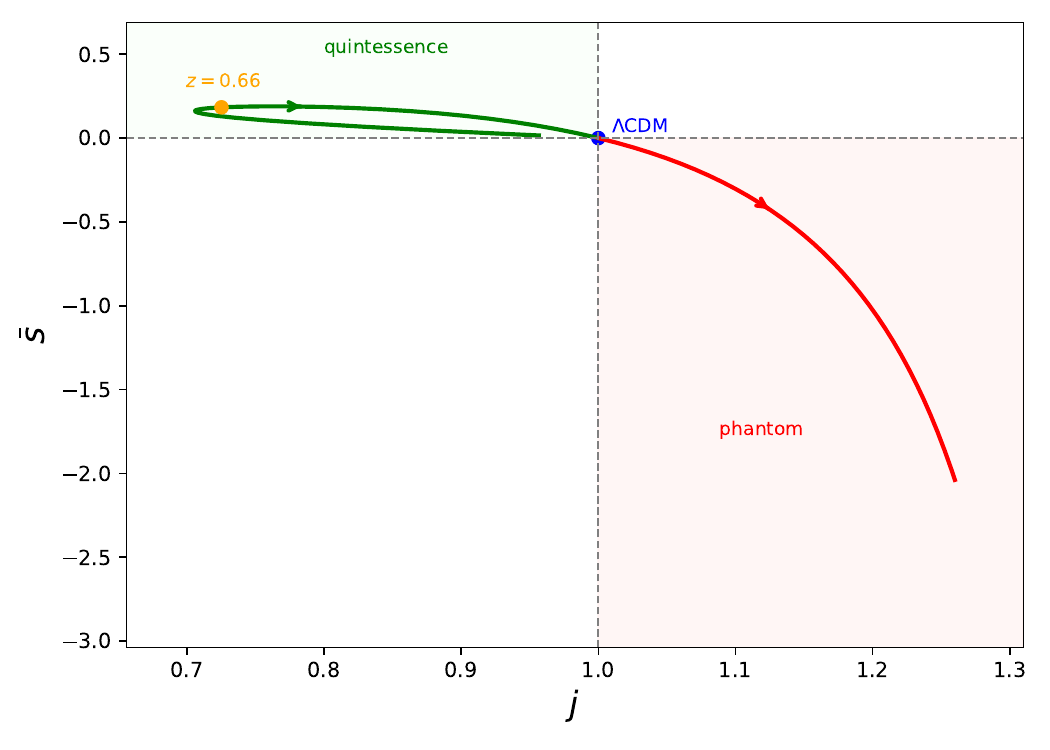}
	\caption{\label{figstatefinder} The behavior of statefinder curve in $\bar{s}-j$ plane for NMHMP gravity model for the best fit values of the parameters as given by table \ref{bestfit}. We have also specified the $\Lambda$CDM point.}
\end{figure}
\begin{align}
	\bar{s}=\frac{2(j-1)}{3(2q-1)}.
\end{align}
For $\Lambda$CDM model, the statefinder pair denotes a point $(1,0)$ in $(j,\bar{s})$ plane. Generally the values $j<1$ and $\bar s>0$ indicates quintessence-like behavior for the dark energy sector but $j>1$ and $\bar s<0$ indicate phantom-like behavior.
In figure \eqref{figstatefinder}, we have plotted the behavior of the statefinder pair for the NMHMP model. It is evident that the DE sector has a quintessence-like behavior at late times, becoming phantom at earlier times. This result is in agreement with our previous analysis of theory using higher derivative of the Hubble function and also the DE equation of state itself. From the statefinder diagram, we can also see that the deceleration to acceleration phase transition occurs in the quintessence epoch of the dark energy.
\subsection{Model comparison}
Let us now compare the results of the NMHMP model with the $\Lambda$CDM model by defining the Bayes factor as
\begin{align}
	B_{N\Lambda} = \frac{\mathcal{Z}_{N}}{\mathcal{Z}_{\Lambda}},
\end{align}
where $\mathcal{Z}_A$ is the marginal likelihood of model $A$. The Bayes factor quantifies how strongly the data favors one model over the other. Here we adopt the Jeffreys scale \cite{jeffreys} in which $|\ln B_{W\Lambda}|<1$ indicated inconclusive evidence, $1<|\ln B_{W\Lambda}|<2.5$ indicated weak evidence, $2.5<|\ln B_{W\Lambda}|<5$ corresponds to moderate evidence and $|\ln B_{W\Lambda}|>5$ indicates strong evidence in favor of the model with higher evidence. With our definitions, positive result favors NMHMP model while negative result favors $\Lambda$CDM model. 
\begin{table}[h!]
	\centering
	\begin{tabular}{|c|c||c|}
		\hline
		Model & $\ln \mathcal{Z}$ & $\ln B_{N\Lambda}$ \\ \hline\hline
		$\Lambda$CDM  & $-900.873\pm 0.331$ & \multirow{2}{*}{$6.508 \pm 0.468$} \\ \cline{1-2}
		NMHMP & $-894.365 \pm 0.331$ & \\ \hline
	\end{tabular}
	\caption{The evidences and Bayes factor for the $\Lambda$CDM and NMHMP models.}
	\label{table3}
\end{table}
In Table~\ref{table3} we have summarized the result of this analysis, for both models. It is evident from these values that the NMHMP model strongly favors over the $\Lambda$CDM model. This means that the data are fitted better in NMHMP model and its predictions would be more reliable than those of $\Lambda$CDM model.

\begin{figure}
	\centering
	\includegraphics[scale=0.7]{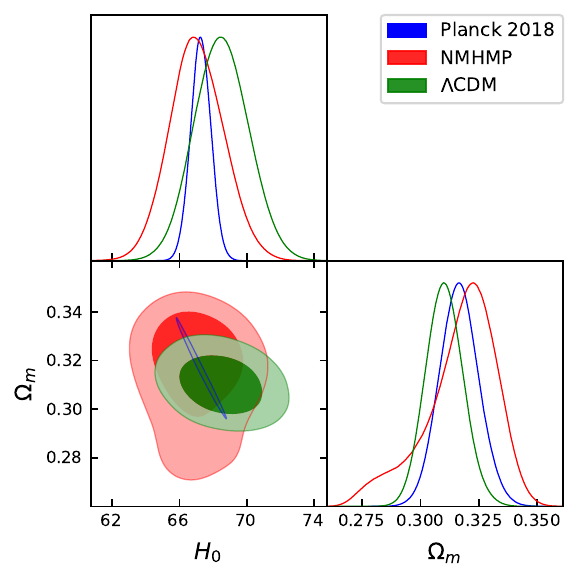}
	\caption{\label{fig2D} The 2D marginalized $H_0$-$\Omega_{m0}$ plot for the Planck data (blue) and for $\Lambda$CDM (green) and NMHMP (red) with $CC+Pantheon^++BAO$ dataset.}
\end{figure}

Let us now consider the tension between the value of the Hubble parameter inferred in this model compared to the value obtained from CMB observations. From the Planck 2018 results \cite{planck} one can read the value of the Hubble parameter as
$$ H_0 = 67.2733^{+0.6002}_{-0.5961}.$$
Now using the value of the Hubble parameter of model from table \ref{bestfit}, one obtains
\begin{table}[h!]
	\centering
	\begin{tabular}{|c||c|c|}
		\hline
		\textbf{Dataset} & \textbf{$H_0$} & \textbf{Tension} \\
		\hline\hline
		Planck 2018& $67.27^{+0.60}_{-0.60}$ & -- \\
		\hline
		NMHMP (CC+PNT+BAO) & $67.04^{+1.60}_{-1.54}$ & $0.14\sigma$ \\
		\hline
		NMHMP (CC+PNT)     & $67.08^{+1.70}_{-1.71}$ & $0.11\sigma$ \\
		\hline
		NMHMP (CC+BAO)     & $69.72^{+2.18}_{-2.19}$ & $1.08\sigma$ \\
		\hline
		$\Lambda$CDM (CC+PNT+BAO) & $68.50^{+1.61}_{-1.61}$ & $0.71\sigma$ \\
		\hline
	\end{tabular}
	\caption{\label{tab:H0_comparison} The tension between Planck 2018 data and the NMHMP model using three difference dataset combinations. We have also included the result for $\Lambda$CDM model.}
\end{table}
One can see from the table that the tension of NMHMP model with the CMB data is about $0.14\sigma$ which indicates that the present model is in full agreement with the Planck observations. It should be noted here that the well-known $H_0$ tension is between the predictions of Planck data and late-time observations with SH0ES calibration \cite{SH0ES}. Here, as we have mentioned before, we have avoided this tension by inferring the value of absolute magnitude $\mathcal{M}$ directly from the cosmological observations, instead of using the SH0ES calibration. As a result, there is no Hubble tension present here.
In figure \eqref{fig2D} we have depicted the 2D marginalized $H_0$-$\Omega_{m0}$ plot for the Planck data \cite{planck}, and for $\Lambda$CDM and NMHMP models with the full dataset $CC+Pantheon^++BAO$. One can see the full agreement of the present model with the Planck 2018 result.

\section{Conclusions and final remarks}\label{sec5}
In this paper, we have considered a non-minimal coupling between matter fields and geometry by assuming an arbitrary function of the Palatini Ricci scalar and the baryonic matter Lagrangian in a hybrid metric-Palatini theory. The hybrid metric-Palatini theory  is build as a modification of the standard general relativity, by inclusion of Palatini independent connection as a new degree of freedom to the existing metric theory. The resulting theory then has two separate parts that can have non-minimal or minimal connection. The first part is made of the metric tensor, mimicking the standard GR theory, and the second part is made of a metric and also an independent connection, which corresponds to the Palatini idea. Earlier ideas couple the matter sector minimally to the the geometry part and assume that the matter sector is made of the metric and matter fields, but not the independent connection. These models are proved to be identical to a (bi-)scalar-tensor theories of gravity. In this paper on the other hand, we have explored the possibility that the matter sector couples non-minimally to the geometry. We have assumed that this coupling allows the matter field to interact with the Palatini Ricci scalar, while leaving the metric sector untouched. Transforming to the scalar-tensor representation, we have seen that the model can be described as a bi-scalar-tensor theory, where one of the scalars dynamically and non-minimally coupled to geometry, mimicking the Brans-Dicke theory, and the other is non-dynamical and couples to the matter sector. Since we have a non-minimal matter coupling in this model, the matter sector is not independently conserved and the evolution of the energy density depends on the behavior of the auxiliary scalar field $\psi$. We have seen however, that the conservation of the energy-momentum tensor can be restored on top of the FRW universe if one assumes the matter Lagrangian to be $L_m=-\rho$. 

It should be noted that since the matter Lagrangian is coupled to the scalar field, the behavior of the matter sector strongly governed by the bi-scalar potential. As we have seen in this paper, the linear choice for the coupling function results in a vanishing potential, which forces the model to exclude the presence of the matter field. The behavior of such a universe is identical to the radiation-dominated universe in standard $\Lambda$CDM theory, where the remaining scalar field plays the role of baryonic matter. However, for more complicated, non-minimal interactions between matter and geometry, one can freely have a matter sector in the theory and the scalar fields can be determined through the scalar field equations of motion.

In this paper, we have considered the cosmological implications of the non-minimal case in details and obtained the model and also the cosmological parameters by fitting the model with three different set of cosmological data. We have used different combinations of the cosmic chronometers, the Pantheon$^+$ dataset and also the DESI DR2 BAO data. In summary the model fits very well with all the datasets and as we have seen, the Jeffreys scale indicates a strong evidence of the present model over the $\Lambda$CDM model. It should be noted that this will not allows us to fully replace the NMHMP model with $\Lambda$CDM, since we have only considered the background evolution of the model. Perturbation analysis of the model, together with other cosmic tests should also be done to decide which model explain the data better. 

The statistical analysis of the model, reveals that the models parameter $n$ has a linear correlation between the parameter $\beta$ and also the abundance $\Omega_{m0}$. As a result, increasing the value of $n$ results in the decrease in $\beta$ and also increase in the matter abundance. This means that larger exponent in the interaction implies more baryonic matter in the universe.

We have seen in this paper that the predictions of the Hubble parameter in NMHMP model is very similar to the $\Lambda$CDM model and the relative difference between the two models is about $2\%$. However, these small differences can be seen from the cosmological functions. In fact, we have seen from the deceleration parameter diagram that the deceleration to acceleration phase transition takes place in a slightly larger redshifts than the $\Lambda$CDM model, implying that the dark energy epoch of the NMHMP model is a little older than its $\Lambda$CDM counterpart. Also, we have seen that the universe experiences smaller amount of late time acceleration in NMHMP model. On the other hand, the DE pressure is a dynamical and increasing function of redshift such that the model predicts a nearly cosmological constant behavior at late times, while the DE dilutes at earlier times. This can also be seen from various DE diagrams we have plotted in this paper, where the late time curves of the NMHMP model approaches the $\Lambda$CDM behavior. This also implies that the effective behavior of the universe is quintessence like in this model mimicking the $\Lambda$CDM model. However, the future behavior in NMHMP is phantom like, showing more dominance of the dark energy at $z<0$. In summary, one can say that the non-minimal coupling between the matter sector and the Palatini part of the geometry can reliably explain the late time behavior of the universe and serves as a good candidate for fitting late-time observational data.


\begin{thebibliography}{99}
	
	\bibitem{acceleration}
	A. G. Riess et al., Astron. J. 116 (1998) 1009.
	
	\bibitem{SCP}
	S. Perlmutter et al., Astrophys. J. 517 (1999) 565.
	
	\bibitem{HZ}
	B. P. Schmidt et al., Astrophys. J. 507 (1998) 46.
	
	\bibitem{DESIDR2}
	M. Abdul-Karim et al. (DESI Collaboration), Phys. Rev. D 112 (2025) 083515.
	
	\bibitem{LCDM}
	P. J. E. Peebles, Astrophys. J. 284 (1984) 439;
	J. P. Ostriker, P. J. Steinhardt, Nature 377 (1995) 600.
	
	\bibitem{CCproblem}
	Y. B. Zeldovich, JETP Lett. 6 (1967) 316;
	S. Weinberg, Rev. Mod. Phys. 61 (1989) 1;
	J. Martin, C. R. Phys. 13 (2012) 566.
	
	\bibitem{conicidence}
	P. J. Steinhardt, in: Critical Problems in Physics, Princeton Univ. Press, Princeton (1997);
	H. E. S. Velten, R. F. vom Marttens, W. Zimdahl, Eur. Phys. J. C 74 (2014) 3160.
	
	\bibitem{hubbletension}
	A. G. Riess et al., Astrophys. J. 826 (2016) 56;
	J. L. Bernal, L. Verde, A. G. Riess, JCAP 10 (2016) 019.
	
	\bibitem{sigmatension}
	C. Heymans et al., Mon. Not. R. Astron. Soc. 432 (2013) 2433;
	H. Hildebrandt et al., Mon. Not. R. Astron. Soc. 465 (2017) 1454.
	
	\bibitem{odint}
	S.~D.~Odintsov, D.~S{\'a}ez-Chill{\'o}n G{\'o}mez and G.~S.~Sharov, Eur. Phys. J. C 85 (2025) 298.
	
	\bibitem{fR}
	H. A. Buchdahl, Mon. Not. R. Astron. Soc. 150 (1970) 1;
	T. P. Sotiriou, V. Faraoni, Rev. Mod. Phys. 82 (2010) 451;
	S. Capozziello, M. De Laurentis, Phys. Rept. 509 (2011) 167;
	S.~Nojiri and S.~D.~Odintsov, Phys. Rept. 505 (2011) 59;
	L. Amendola, S. Tsujikawa, Dark Energy: Theory and Observations, Cambridge Univ. Press (2010).
	
	\bibitem{higher}
	L. Randall, R. Sundrum, Phys. Rev. Lett. 83 (1999) 4690;
	L. Randall, R. Sundrum, Phys. Rev. Lett. 83 (1999) 3370;
	N. Arkani-Hamed, S. Dimopoulos, G. Dvali, Phys. Lett. B 429 (1998) 263.
	
	\bibitem{massive}
	M. Fierz, W. Pauli, Proc. R. Soc. Lond. A 173 (1939) 211;
	S. F. Hassan, R. A. Rosen, A. Schmidt-May, JHEP 2012 (2012) 026;
	C. de Rham, G. Gabadadze, A. J. Tolley, Phys. Rev. Lett. 106 (2011) 231101;
	S. Shahidi, in: Proc. 14th Marcel Grossmann Meeting (2017);
	Z. Haghani, H. R. Sepangi, S. Shahidi, Phys. Rev. D 87 (2013) 124014;
	N. Khosravi, H. R. Sepangi, S. Shahidi, Phys. Rev. D 86 (2012) 043517;
	N. Khosravi, N. Rahmanpour, H. R. Sepangi, S. Shahidi, Phys. Rev. D 85 (2012) 024049.
	
	\bibitem{WCF}
	H. Weyl, Sitzungsber. Preuss. Akad. Wiss. Berlin (Math. Phys.) 1918 (1918) 465;
	E. Cartan, C. R. Acad. Sci. (Paris) 174 (1922) 593;
	P. Finsler, Über Kurven und Flächen in allgemeinen Räumen, Dissertation, Göttingen (1918);
	H. Weyl, Ann. Phys. (Leipzig) 360 (1918) 117;
	T. Harko, S. Shahidi, Eur. Phys. J. C 84 (2024) 509;
	D.–I. Visa, T. Harko, S. Shahidi, Phys. Dark Univ. 46 (2024) 101720;
	J.–Z. Yang, S. Shahidi, T. Harko, Eur. Phys. J. C 82 (2022) 1171;
	R. Hama, T. Harko, S. V. Sabau, S. Shahidi, Eur. Phys. J. C 81 (2021) 742;
	A. Bouali et al., Eur. Phys. J. C 83 (2023) 121;
	R. Hama, T. Harko, S. V. Sabau, Eur. Phys. J. C 83 (2023) 1030;
	D. Puetzfeld, R. Tresguerres, Class. Quantum Grav. 18 (2001) 677;
    Z. Haghani, N. Khosravi, S. Shahidi, Class. Quantum Grav. 32 (2015) 215016;
	Z. Haghani, T. Harko, H. R. Sepangi, S. Shahidi, JCAP 10 (2012) 061;
	Z. Haghani, T. Harko, H. R. Sepangi, S. Shahidi, Phys. Rev. D 88 (2013) 044024.

	\bibitem{ghost}
	D. G. Boulware, S. Deser, Phys. Rev. D 6 (1972) 3368;
	M. Ostrogradsky, Mem. Acad. St. Petersbourg 6 (1850) 385;
	A. Klein, D. Roest, JHEP 07 (2016) 130.
	
	\bibitem{scalarvector}
	J. W. Moffat, JCAP 2006 (2006) 004;
	G. W. Horndeski, Int. J. Theor. Phys. 10 (1974) 363;
	J. D. Bekenstein, Phys. Rev. D 70 (2004) 083509;
	T. Kobayashi, Rep. Prog. Phys. 82 (2019) 086901;
	A. Nicolis, R. Rattazzi, E. Trincherini, Phys. Rev. D 79 (2009) 064036;
	C. Deffayet et al., Phys. Rev. D 84 (2011) 064039;
	G. Tasinato, JHEP 2014 (2014) 067;
	L. Heisenberg, JCAP 2014 (2014) 015.
	
	\bibitem{modifiedmatter}
	N. Katirci, M. Kavuk, Eur. Phys. J. Plus 129 (2014) 163;
	R. A. C. Cipriano et al., Universe 10 (2024) 339;
	R. Ferraro, F. Fiorini, Phys. Rev. D 75 (2007) 084031;
	M. Roshan, F. Shojai, Phys. Rev. D 94 (2016) 044002;
	Ö. Akarsu, N. Katirci, S. Kumar, PoS CORFU2017 (2018) 105.
	
	\bibitem{derivativematter}
	Z. Haghani, S. Shahidi, Phys. Dark Univ. 30 (2020) 100683;
	P. Asimakis et al., Phys. Rev. D 107 (2023) 104006;
	Z. Haghani, S. Shahidi, Eur. Phys. J. Plus 135 (2020) 509.
	
	\bibitem{secondderivative}
	Z. Haghani, T. Harko, S. Shahidi, Phys. Dark Univ. 44 (2024) 101448.
	
	\bibitem{fRT}
	T. Harko, F. S. N. Lobo, S. Nojiri, S. D. Odintsov, Phys. Rev. D 84 (2011) 024020;
	T. Harko, F. S. N. Lobo, Eur. Phys. J. C 70 (2010) 373;
	Z. Haghani et al., Phys. Rev. D 88 (2013) 044023;
	Z. Haghani, T. Harko, Eur. Phys. J. C 81 (2021) 615;
	T. B. Gonçalves, J. L. Rosa, F. S. N. Lobo, Phys. Rev. D 109 (2024) 084008;
	T. Harko, F. S. N. Lobo, Int. J. Mod. Phys. D 29 (2020) 2030008.
	
	\bibitem{debate}
	O. Lacombe, S. Mukohyama, J. Seitz, JCAP 05 (2024) 064.
	
	\bibitem{answer}
	T. Harko, M. A. S. Pinto, S. Shahidi, Phys. Dark Univ. 48 (2025) 101863.
	
	\bibitem{mattercreation}
	I. Prigogine, J. Geheniau, E. Gunzig, P. Nardone, Gen. Rel. Grav. 21 (1989) 767;
	T. Harko, F. S. N. Lobo, J. P. Mimoso, D. Pavón, Eur. Phys. J. C 75 (2015) 386.
	
	\bibitem{palatini}
	A. Palatini, Rend. Circ. Mat. Palermo 43 (1919) 203;
	A. Delhom, D. Rubiera-Garcia, in: Modified Gravity and Cosmology, Springer, Cham (2021).
	
	\bibitem{palatinifR}
	D. N. Vollick, Phys. Rev. D 68 (2003) 063510;
	G. J. Olmo, Int. J. Mod. Phys. D 20 (2011) 413.
	
	\bibitem{hybrid}
	T. Harko et al., Phys. Rev. D 85 (2012) 084016;
	S. Capozziello et al., Universe 1 (2015) 199;
	N. Tamanini, C. G. Boehmer, Phys. Rev. D 87 (2013) 084031;
	J. L. Rosa, S. Carloni, J. P. S. Lemos, F. S. N. Lobo, Phys. Rev. D 95  (2017) 124035;
	J. L. Rosa, S. Carloni, J. P. S. Lemos, Phys. Rev. D 101  (2020) 104056;
	R. Aliannejadi, Z. Haghani, Iran. J. Astron. Astrophys. 4 (2024) 183.
	
	\bibitem{BD}
	C. Brans, R. H. Dicke, Phys. Rev. 124 (1961) 925;
	M. S. Berman, Int. J. Theor. Phys. 29 (1990) 571.
	
	\bibitem{coshybrid}
	B. Asfour et al., Phys. Lett. B 853 (2024) 138679;
	J. L. Rosa, Eur. Phys. J. C 84 (2024) 895;
	I. D. Gialamas et al., Int. J. Geom. Meth. Mod. Phys. 20 (2023) 2330007;
	M. He, Y. Mikura, Y. Tada, JCAP 05 (2023) 047;
	S. Capozziello et al., JCAP 04 (2013) 011;
	F. Bombacigno, F. Moretti, G. Montani, Phys. Rev. D 100 (2019) 124036;
	S. Shahidi, S. Kayedi, Eur. Phys. J. C 85 (2025) 721.
	
	\bibitem{chameleon}
	J. Khoury, A. Weltman, Phys. Rev. Lett. 93 (2004) 171104;
	J. Khoury, A. Weltman, Phys. Rev. D 69 (2004) 044026.
	
	\bibitem{matterPalatini}
	T. Harko, T. S. Koivisto, F. S. N. Lobo, Mod. Phys. Lett. A 26 (2011) 1467.
	
	\bibitem{diffeomorphisminvariance}
	E. Noether, Nachr. Ges. Wiss. Göttingen Math.-Phys. Kl. 1918 (1918) 235;
	R. Dick, Int. J. Theor. Phys. 32 (1993) 109.
	
	\bibitem{CCdata}
	E. K. Li et al., Mon. Not. R. Astron. Soc. 501 (2021) 4452;
	D. Stern et al., JCAP 02 (2010) 008;
	M. Moresco et al., JCAP 08 (2012) 006;
	M. Moresco et al., JCAP 05 (2016) 014;
	M. Moresco, Mon. Not. R. Astron. Soc. 450 (2015) L16;
	A. L. Ratsimbazafy et al., Mon. Not. R. Astron. Soc. 467 (2017) 3239;
	C. Zhang et al., Res. Astron. Astrophys. 14 (2014) 1221;
	H. Singirikonda, S. Desai, Eur. Phys. J. C 80 (2020) 694.
	
	\bibitem{popCC}
	M. Moresco et al., Astrophys. J. 898 (2020) 82.
	
	\bibitem{PANdata}
	D. Brout et al., Astrophys. J. 938 (2022) 110.
	
		\bibitem{SH0ES}
	A. G. Riess et al. (SH0ES Collaboration), Astrophys. J. Lett. 934 (2022) L7.
	
	\bibitem{statefinder}
	V. Sahni, T. D. Saini, A. A. Starobinsky, U. Alam, JETP Lett. 77 (2003) 201.
	
	\bibitem{jeffreys}
	H. Jeffreys, Theory of Probability, 3rd ed., Oxford Univ. Press, Oxford (1961).
	
	\bibitem{planck}
	N. Aghanim et al. (Planck Collaboration), Astron. Astrophys. 641 (2020) A6.
	
\end{thebibliography}
\end{document}